\documentclass[twocolumn,tighten,preprint,times]{aastex631}

\usepackage{bm}
\usepackage[utf8]{inputenc}
\usepackage{amsmath}
\usepackage{graphicx}
\usepackage{appendix}
\usepackage{xcolor}
\usepackage{booktabs}
\usepackage{subfigure}
\usepackage{multirow}

\newcommand\cebm{\textrm{CEBM}}

\newcommand\Mpc{\mathrm{Mpc}}
\newcommand\kpc{\mathrm{kpc}}
\newcommand\Msun{M_{\sun}}
\newcommand\km{\mathrm{km}}
\newcommand\yr{\mathrm{yr}}
\newcommand\sm{\mathrm{s}}

\newcommand\learningrate{\mathcal{R}_l}
\newcommand\binning{\mathcal{B}}
\newcommand\maxbins{\mathcal{Q}_\mathrm{max}}
\newcommand\maxbinstwod{\mathcal{Q}_\mathrm{max,2D}}

\newcommand\betay{\beta_y}
\newcommand\fyi{f_y^i}
\newcommand\fyij{f_y^{ij}}
\newcommand\fii{f^i}
\newcommand\fij{f^{ij}}

\newcommand\vtheta{\bm{\theta}}
\newcommand\thetai{\theta_{i}\!}

\newcommand\np{n_\mathrm{p}} 
\newcommand\mstar{M_{\star}} 
\newcommand\sfr{S\!F\!R} 
\newcommand\mvir{M_{\mathrm{vir}}} 
\newcommand\vpeak{v_\mathrm{peak}} 
\newcommand\rhoR{\rho_1} 
\newcommand\TR{T_1} 
\newcommand\upR{\Upsilon_{0.1}} 

\newcommand\ebmy{\gamma(y|\vtheta)}
\newcommand\ebmsfr{\gamma(\sfr|\vtheta)}
\newcommand\ebmms{\gamma(\mstar|\vtheta)}

\newcommand\basesfr{\gamma(\sfr|\vtheta')}
\newcommand\basems{\gamma(\mstar|\vtheta')}
\newcommand\basey{\gamma(y|\vtheta')}

\newcommand\cebmsfr{\Gamma(\sfr|\vtheta')}
\newcommand\cebmms{\Gamma(\mstar|\vtheta')}
\newcommand\cebmy{\Gamma(\mstar|\vtheta')}

\newcommand\ebmsfrp{\gamma(\sfr|\vtheta')}
\newcommand\ebmmsp{\gamma(\mstar|\vtheta')}

\newcommand\outsfr{\delta(\sfr|\vtheta')}
\newcommand\outms{\delta(\mstar|\vtheta')}
\newcommand\outy{\delta(y|\vtheta')}

\newcommand\phisfr{\phi_{\sfr}(\vtheta')}
\newcommand\phims{\phi_{\mstar}(\vtheta')}
\newcommand\phiy{\phi_{y}(\vtheta')}

\newcommand\classsfr{\phisfr}
\newcommand\classms{\phims}
\newcommand\classy{\phiy}

\newcommand\cfave{\bar{f}}
\newcommand\cfeat{\tilde{f}}

\submitjournal{AAS Journals}

\begin{document}

\title{Revealing the Galaxy-Halo Connection Through Machine Learning}
\shorttitle{Machine Learning the Galaxy-Halo Connection}

\correspondingauthor{}
\email{rhausen@ucsc.edu, brant@ucsc.edu}

\author[0000-0002-8543-761X]{Ryan Hausen}
\affiliation{Department of Computer Science and Engineering, University of
             California, Santa Cruz, 1156 High Street, Santa Cruz, CA 95064 USA}

\author[0000-0002-4271-0364]{Brant E. Robertson}
\affiliation{Department of Astronomy and Astrophysics, University of California,
             Santa Cruz, 1156 High Street, Santa Cruz, CA 95064 USA}

\author[0000-0003-0861-0922]{Hanjue Zhu}
\affiliation{Department of Astronomy and Astrophysics, University of Chicago,
             5640 S. Ellis Ave, Chicago, IL 60637 USA}

\author[0000-0001-5925-4580]{Nickolay Y.\ Gnedin}
\affiliation{Fermi National Accelerator Laboratory,
Batavia, IL 60510, USA}
\affiliation{Kavli Institute for Cosmological Physics,
The University of Chicago}
\affiliation{Department of Astronomy and Astrophysics, University of Chicago,
             5640 S. Ellis Ave, Chicago, IL 60637 USA}

\author[0000-0002-6336-3293]{Piero Madau}
\affiliation{Department of Astronomy and Astrophysics, University of California,
             Santa Cruz, 1156 High Street, Santa Cruz, CA 95064 USA}

\author[0000-0001-9735-7484]{Evan E. Schneider}
\affiliation{Department of Physics and Astronomy, University of Pittsburgh,
             100 Allen Hall 3941 O'Hara St., Pittsburgh, PA 15260 USA}

\author[0000-0002-7460-8129]{Bruno Villasenor}
\affiliation{Department of Astronomy and Astrophysics, University of California,
             Santa Cruz, 1156 High Street, Santa Cruz, CA 95064 USA}

\author[0000-0003-4761-2197]{Nicole E. Drakos}
\affiliation{Department of Astronomy and Astrophysics, University of California,
             Santa Cruz, 1156 High Street, Santa Cruz, CA 95064 USA}

\begin{abstract}
Understanding the connections between galaxy stellar mass, star formation rate,
and dark matter halo mass represents a key goal of the theory of galaxy
formation. Cosmological simulations that include hydrodynamics, physical
treatments of star formation, feedback from supernovae, and the radiative
transfer of ionizing photons can capture the processes relevant for establishing these connections. The complexity of these
physics can prove difficult to disentangle and
obfuscate how mass-dependent trends in the
galaxy population originate. Here, we train a machine learning method
called Explainable Boosting Machines (EBMs) to infer how the
stellar mass and star formation rate
of nearly 6 million galaxies simulated by the Cosmic Reionization on
Computers (CROC) project depend on the
physical properties of halo mass, the peak circular velocity
of the galaxy during its formation history $\vpeak$, cosmic environment,
and redshift. The resulting EBM models reveal the relative
importance of these properties in setting galaxy stellar mass
and star formation rate, with $\vpeak$ providing the most
dominant contribution.
Environmental properties provide substantial improvements
for modeling the stellar mass and star formation rate in only
$\lesssim10\%$ of the simulated galaxies.
We also provide alternative formulations
of EBM models that enable low-resolution simulations, which
cannot track the interior structure of dark matter halos, to
predict the stellar mass and star formation rate of
galaxies computed by high-resolution simulations with detailed
baryonic physics.
\end{abstract}

\section{Introduction}
\label{sec:Introduction}

Numerical simulation enables theoretical
models of galaxy formation to include
detailed physical models for baryonic
processes. Simulations can capture
the physics of cooling,
supernova feedback, radiative feedback
and ionization, and the role of
dynamics simultaneously while
tracking the growth of cosmological
structure formation
\citep[e.g.,][]{schaye2015a,pillepich2018a,dave2019a}.
The simulated
galaxy populations that result
from these models reproduce
observed stellar mass sequences
such as the main sequence of
star-forming galaxies
\citep{brinchmann2004a,noeske2007a}
or the red sequence of quiescent
galaxies \citep{faber2007a}.
The quest for realism in modeling
these observed trends has also
added substantial complexity, such
that understanding which physical
properties of a galaxy most influence
its stellar mass and star formation
rate can prove challenging.
Many theoretical frameworks to describe
these
relations have been developed \citep[e.g.,][]{wechsler2018a}, including
halo occupation distribution models
\citep[e.g.,][]{jing1998a}, subhalo abundance
matching \citep{vale2004a,conroy2006a}, and
semi-analytic models \citep[for a review, see][]{somerville2015a}.
The complex physics encoded by these models
and simulations can be difficult to
interpret, and the relative contribution of
baryonic feedback, dark matter halo formation,
and environment in setting galaxy properties
remains challenging to disentangle.

This complexity extends to cosmological
models of galaxy formation in the reionization epoch.
To capture the
distribution of sizes of ionized regions with converged simulations \citep{Iliev2014} and the largest observed features, such as dark gaps \citep{zhu2021a},
the volume of reionization simulations should extend
to a least several hundred megaparsecs.
Modeling such large volumes in a single simulation
while maintaining the spatial
resolution needed to include the complex physics
of the current state-of-the-art projects, such as
Cosmic Reionization on
Computers \citep[CROC][]{gnedin2014a}, THESAN \citep{thesan1}, or Cosmic Dawn \citep[CoDa,][]{coda1,coda2}, remains computationally infeasible.
Instead, we desire an intermediate approach where large volumes are
simulated and the physics of galaxy formation are implemented
with a approximate model that recovers the mean trends for
galaxy baryonic properties predicted by more detailed calculations.
With this goal in mind, a model for reionization sources that
encapsulates the results of projects like CROC in a simple module
is the first necessary step for deploying lower resolution
simulations with much larger ($L\sim 500 {\rm cMpc}$) simulation volumes.
If the stellar mass and star formation rates
of ionizing sources can be predicted from their
dark matter halo properties and environment, then we
can account for the ionizing photons produced by these
sources in large-box simulations
of the reionization process without resolving the
baryonic physics in detail.

This work employs a machine learning method
called Explainable Boosting Machines
\citep[EBMs][]{Lou2013} to infer how
stellar mass $\mstar$ and star formation rate
$\sfr$
depend on the physical parameters $\vtheta$ of
a host galaxy.
In this work, we use the
galaxy populations from the
CROC simulations
to provide our training and test data
that populate samples in the multidimensional
parameter space of $\mstar$-$\sfr$-$\vtheta$.
For the additional parameters $\vtheta$, we use a wide
range of physical characteristics measured for
galaxies in CROC including the virial
mass $\mvir$, redshift $z$, environmental
properties
averaged on a length scale $R$, and the
maximum peak circular velocity $\vpeak$.
We can then use this approximate machine
learning-based EBM model for galaxy formation
as a basis for future development to
incorporate the CROC galaxy population as
sources in lower resolution, large-volume
reionization simulations.

EBMs represent a form of
Generalized Additive Models \citep[][GAMs]{Hasti1986} where the
dependencies of a \emph{target}
quantity, such as $\mstar$ or $\sfr$,
on each physical
parameter $\theta_i$ are encapsulated
by
\emph{feature functions} of one parameter (e.g., $\fii(\theta_i)$)
or \emph{interaction functions} of two parameters (e.g., $\fij(\theta_i,\theta_j)$).
An EBM model is trained to fit these
functions from a provided multidimensional dataset.
The predicted value of the target quantity
given the parameters (e.g., $\ebmms$) is then
a sum of the functions $\fii$
and $\fij$. EBM models are often described
as \emph{interpretable} because the magnitudes of the
functions $\fii$ and $\fij$ directly
indicate the relative importance
of $\vtheta$ in determining the target
quantity. If a given parameter $\thetai$
is unimportant for determining the target
quantity, the EBM will find $\fii\!\to\!0$.
A formal defintion of the EBM is provided
in Section \ref{sec:ebm}.

Previous works have applied machine learning models
to infer connections between simulated galaxy properties.
\citet{lovell2021a} use a tree-based learning method called Extremely Randomized
Trees to map baryon information to dark matter halos in the EAGLE simulations.
\citet{xu2021a} train a
Random Forest to predict the number of central and satellite galaxies in dark
matter halos in the Millennium simulation.
\citet{machado2021a} used an XGBoost model to predict gas shapes
in dark matter halos in the IllustrisTNG simulations.
\citet{bluck2022a} used Random Forest classifiers to study quenching
mechanisms in observations, semianalytical models, and cosmological
simulations. \citet{piotrowska2022a} also used Random Forest classifiers
to examine how supermassive black hole feedback quenches central
galaxies in the EAGLE, Illustris, and  IllustrisTNG simulations.
Our approach complements these prior works by studying
the detailed connection between halo and environmental
properties, star formation rate, and stellar mass in
a model that can be directly implemented in future
large-volume cosmological simulations with limited
spatial resolution.

The paper is organized as follows. In \S \ref{sec:methods} we review the
EBM methodology, define our training dataset and procedure, and introduce
the evaluation metrics used to assess the performance of the model.
In \S \ref{sec:results} we present the average contribution of each
parameter to the target quantities, the best-fit feature and interaction
functions, and the performance of the model in determining the
distributions of
stellar mass and star formation rate as a function of halo virial mass.
We then explore in \S \ref{sec:cebm} methods for constructing
\emph{composite}
EBM models to recover the stellar mass and star formation rate of simulated
galaxies that only use instantaneous halo virial properties and environmental
measures (i.e., excluding $\vpeak$). We discuss our results in \S \ref{sec:discussion},
and summarize them and conclude in \S \ref{sec:summary}. The Appendicies of the
paper provide detailed model results for the EBM for $\mstar$ (\S \ref{appendix:ebm_mstar}),
the mathematical formalism of the composite EBM model (\S \ref{appendix:cebm}),
and detailed composite EBM model results for star formation rate (\S \ref{appendix:cebm_sfr})
and stellar mass (\S \ref{appendix:cebm_mstar}).

\section{Methods}
\label{sec:methods}

To infer the connection between $\mstar$, $\sfr$, and other physical properties
of simulated galaxies, we apply EBM models to the CROC simulated galaxy
catalogs. In \S \ref{sec:ebm}, we define the EBM model.
We select our model parameters and describe the
simulated
galaxy catalog used to train the model in \S \ref{sec:dataset}.
The training procedure is outlined
in \S \ref{sec:training}.

\subsection{Explainable Boosting Machines}
\label{sec:ebm}

Explainable Boosting Machine \citep[][EBM]{Lou2013} models
provide a fitted representation of the relationship
between the target quantities $y$ and the parameters
$\vtheta$. EBMs are an extension of
Generalized Additive Models \citep[][GAMs]{Hasti1986},
which represent target quantities $y$ as the sum
of learned univariate
functions $\fii(\theta_i)$ that
depend on only one parameter $\theta_i$. EBMs
extend GAMs by including both univariate
functions $\fii(\theta_i)$
and bivariate functions $\fij(\theta_i,\theta_j)$
that represent dependencies on pairs of
features $(\theta_i,\theta_j)$ beyond the
dependence of the target quantity
on either feature independently.
Both EBMs and GAMs are forms of
regression where the feature functions $\fii$
and $\fij$ can be quite general.

The EBM aims to encode the average dependence
of a target quantity $y$ on the parameters
$\vtheta$.
Mathematically, an EBM can therefore
be represented as
\begin{equation}
\label{eqn:ebm}
\ebmy = \betay + \sum_{i=0}^{\np-1} \fyi(\theta_i) + \sum_{i=0,i \neq j}^{\np-1}\sum_{j=0}^{\np-1}  \fyij(\theta_i, \theta_j)
\end{equation}
\noindent
where $\ebmy$ is the predicted value of the target quantity $y$ given $\np$
parameters $\vtheta \in \mathbb{R}^{n}$ from the dataset.
We will refer to learned parameter $\betay$ as the \emph{baseline value} of the
target quanity $y$.
Though $\fyi$ and $\fyij$
can be any interpretable function (e.g., linear regression, splines, etc.),
\citet{Lou2012} found that gradient boosted trees \citep{Friedman2001} work best
in practice. Using gradient boosted trees, the functions $\fyi$ and $\fyij$
will be piece-wise one- and two-dimensional functions, respectively.
By expressing the dependence of $y$ on $\vtheta$ directly
through the functions $\fyi$ and $\fyij$, EBMs are interpretable
and decomposeable.
Further, after training is complete the learned tree-based
functions $\fyi$ and $\fyij$
can be formulated as look-up tables for
performant inference.

\subsection{Simulated Galaxy Catalog Training Set}
\label{sec:dataset}

To engineer an EBM that describes the connection between
simulated galaxy properties, their host dark matter halos,
and features of the extrinsic environment, we turn to
established observations and theoretical modeling to
inform our choices for constructing a training dataset.

The stellar--mass---halo--mass (SHMR) has been directly constrained out to redshifts $z\lesssim0.05$ and galaxy masses $M_{\rm vir}>10^{12} M_\odot$ using galaxy kinematics \citep[e.g.][]{more2009a,li2012a}, $X$-ray observations \citep[e.g][]{lin2004a,kravstov2018a} and gravitational lensing \citep[e.g.][]{mandelbaum2005a,velander2014a}. These constraints can be extended to higher redshifts ($z<10$) and lower masses ($M_{\rm vir}<10^{10}$) by including halo--galaxy connection modeling \citep[e.g.][]{nelson2015a,croton2016a,rodriguez2017a,behroozi2019a,girelli2020a}. Such models consistently
infer that the average stellar mass of galaxies
increases with halo mass.

At fixed redshift and halo mass, average galaxy masses of central galaxies differ from satellite galaxies. Halos grow through hierarchical merging, in which small halos merge to form larger halos. As subhalos merge into larger halos, tidal heating and stripping reduce the mass of the more extended dark matter halo, while the satellite galaxy mass remains largely unaffected. For this reason, galaxy mass often correlates better with halo properties at the time of accretion than the current halo mass \citep[e.g.][]{conroy2006a,vale2006a, moster2010a, reddick2013a}. In particular, SHAM models find that using the halo peak circular velocity, $v_{\rm peak}$, to assign galaxy masss and/or luminosity best reproduces observed galaxy clustering \citep[e.g.][]{reddick2013a,hearin2013a,lehmann2017a}.

Star formation rates correlate tightly with galaxy masses, and increase with redshift at fixed stellar mass
\citep[e.g.][]{noeske2007a,stark2009a,bouwens2012a}. While these trends hold on average, there is a distinct bimodal distribution in the star formation rates of galaxies, corresponding to star-forming and quiescent populations \citep[e.g.][]{balogh2004a}. The observed fraction of quiescent galaxies increases as the Universe evolves \citep[e.g.][]{tomczak2014a}, with the interpretation that some mechanism turns off star formation in galaxies.
Many quenching mechanisms have been proposed, including secular/mass quenching \citep[e.g.][]{kauffmann2004a,contini2020a} and environmental quenching \citep[e.g.][]{davies2016a, trussler2020a}.
Which of these processes dominate may vary with redshift \citep{kalita2021a}.

Overdense environments may cause environmental quenching,
by providing close pairs that can suppress gas accretion (``strangulation"), removing gas through ram-pressure stripping, or disrupting by interactions with other galaxies (``harassment"). Environment thereby
influences star formation rates, and
low-mass satellite galaxies are typically the most
prone to environmental quenching  \citep[e.g.][]{davies2019a}.

Given these established trends, galaxy mass and star formation rate may
depend on redshift, halo mass, peak circular velocity,
and environmental properties.
We will therefore
select corresponding parameters from the CROC simulated galaxy catalogs
to provide our dataset
for training the EBM models. Details of the
CROC simulations can be found in \cite{gnedin2014a}.
At a range of redshifts $z$ during the simulation,
the computational grid and particle properties are written
to disk. These simulation snapshots are post-processed
to identify virialized galaxies, as described in \citet{zhu2020a},
and the properties of the simulated galaxies
are recorded in catalogs. Merger trees are used
to identify the properties of simulated galaxies across
redshift.

For our target quantities $y$,
in this work we will model stellar mass
$\mstar$ [$h^{-1}\Msun$] and star formation
rate $\sfr$ [$\Msun~\yr^{-1}$].
The parameters $\vtheta$ selected
from the simulated catalog include both intrinsic
properties of galaxies and extrinsic properties
set by the large scale environment.
For intrinsic properties we include the
galaxy virial mass $\mvir$ [$h^{-1}\Msun$], the
redshift $z$ at which the simulated galaxy properties were measured,
and the maximum peak circular velocity
$\vpeak$ [$\km~\sm^{-1}$] measured over the formation history
of each galaxy.
The extrinsic properties used are defined by
a length scale $R$ measured relative to each simulated
galaxy. We follow convention and substitute $R$ with a
numerical value that indicates a number of comoving Mpc (e.g.,
$\sigma_8$ is the rms density fluctuations measured in
spheres of radius of $R=8\Mpc$).
We compute an environmental density
$\rhoR \equiv 1 +\Delta_1$, where $\Delta_1$ is the dimensionless
matter overdensity measured within 1 $\Mpc$.
We include an environmental gas temperature
$\TR$ [K] averaged on 1 $\Mpc$ scales. From
each simulated galaxy we also find the virial
mass $M_{\mathrm{max}, 0.1}$
of the most
massive neighboring halo within 100 $\kpc$.
We then define the mass ratio
$\upR \equiv 1 + M_{\mathrm{max}, 0.1}/\mvir$

The simulated galaxy catalogs include
roughly 8,426,327 objects covering a wide
range of halo masses, stellar masses,
star formation rates, redshifts, and
other extrinsic properties.
From the catalog of simulated galaxies, objects with
a $\sfr<0.001$ $\Msun~\yr^{-1}$
were excluded owing to resolution effects artificially
limiting their star formation rates.
After this culling, the catalog contained
5,950,357 objects that formed our dataset.
At this stage, we constructed the training and test
datasets from our catalog
using the parameter vector $\vtheta = [\mvir, z, \vpeak, \rhoR, \TR, \upR]$
to model the target quantities $\bm{y} = [\mstar, \sfr]$.
We use $k$-fold cross-validation \citep{hastie2001} with
$k=5$, such that the test/training split is 20\%/80\% for
each $k$-folding.

\subsection{Training Procedure}
\label{sec:training}

The calculations presented in this paper
leverage the InterpretML \citep{nori2019}
implementation of EBMs, using the hyperparameters in Table
\ref{table:train_params}. These InterpretML
hyperparameters control the number of bins in the piece-wise
$\fyi$ and $\fyij$ functions ($\maxbins$, $\maxbinstwod$),
the distribution of bins across the fitted domain ($\binning$),
and the learning rate of the optimization scheme
($\learningrate$).
The \citet{nori2019} implementation trains an EBM in two
phases. First, the univariate functions are optimized using a gradient boosting
approach applied round-robin on each parameter, as
detailed in \citet{Lou2012}. After the univariate functions have converged, the
interaction terms are computed and the bivariate functions are optimized
according the GA2M/FAST algorithms detailed in \citet{Lou2013}.
During training we use $k$-fold cross-validation, and merge the
training and test datasets for the final performance evaluation of the model.

We evaluate the EBM performance using the mean absolute error (MAE), a variance
metric $r^2$, and the total
outlier fraction $\zeta_k$. These statistics provide measures of how well the EBM
reproduces the mean trends in the target quantities $y$ as a function of the
features $\theta$, the width of the distribution about the mean trends
in the training data, and the tails of that distribution.

We calculate the MAE of the model applied to the simulated
galaxy sample as

\begin{equation}
    \label{eqn:mean_error}
    \textrm{MAE} = \frac{1}{N}\sum_{i=0}^{N-1} |y_i - \hat{y}_i|,
\end{equation}
\noindent
where $N$ is the number of objects, $y_i$ is the true value of the target quantity
for object $i$,
and
$\hat{y}_i$ is the predicted value from the model for object $i$.

We compute the $r^2\in[0,1]$ variance metric as
\begin{equation}
    \label{eqn:r^2}
    r^2 = 1 - \frac{\sum_{i=0}^{N-1} (y_i - \hat{y}_i)^2}{\sum_{i=0}^{N-1} (y_i - \overline{y})^2},
\end{equation}
\noindent
which provides a measure of how well the model captures the variance in the data
relative to the mean $\overline{y}$,
with $r^2=1$ reflecting a perfect reproduction of the distribution
of $y$ in the training
dataset.
Note that the
feature and interaction functions $\fyi$ and $\fyij$
have a finite range, and thus not all values $y_i$
can be represented by Equation \ref{eqn:ebm} even when the input
parameters $\vtheta$ vary about the mean
trends with halo mass or environment. Hence, even for
high quality EBM models $r^2<1$ and we expect outliers.
The $\zeta_k$ metric represents the
fraction of the total dataset that lies
outside the range of predicted values,
$\{\hat{y}\}$, as a function of one of the features $\theta_k$.
We define
\begin{equation}
    \label{eqn:outlier_fraction}
    \zeta_k = \frac{1}{N}\sum_{i=0}^{N-1} g_{k,i}(y_i, \theta_{k,i})
\end{equation}
\noindent
where the index $i$ runs over the total number of samples $N$
and $g_{k,i}(y_i, \theta_{k,i})$ is a
function that returns $1$ if the true target quantity for object $i$ lies
outside the predicted range, i.e.,
$y_i\not\in\{\hat{y}\}$.
In practice, we compute the outlier fraction for feature $k=\log_{10}\mvir$,
and use
2D histograms of ($y_i$, $\theta_{k,i}$) and ($\hat{y}_i$, $\theta_{k,i}$)
to calculate $g_{k,i}$.

In Table \ref{table:ebm_train_results}
we present the evaluation metrics for our
EBM model fully trained on the simulated
galaxy catalog.
For the EBM model for star
formation rate ($y=\sfr$), we find
a $\mathrm{MAE}\sim0.14 \log_{10} \Msun~\yr^{-1}$, a variance metric
$r^2\sim0.9$,
and an outlier fraction of $<3\%$.
For the EBM model for stellar mass
($y=\mstar$),
we report a $\mathrm{MAE}\sim0.19 \log_{10} \Msun~\yr^{-1}$,
a variance metric
$r^2\sim0.88$, and an outlier fraction of $<1\%$.
The good performance of the EBM models in these
metrics reflects the abililty of the EBMs to
capture both the mean trends and full distributions
of the target quantities $y=[\mstar,\sfr]$ in the
training set given
the parameters
$\vtheta = [\mvir, z, \vpeak, \rhoR, \TR, \upR]$.
We describe the detailed model results in
\S \ref{sec:results}.

\begin{table}
    \centering
    \begin{tabular}{c c}
    \multicolumn{2}{c}{EBM Training Hyperparameters} \\
    \toprule
    Hyperparameter & Value \\
    \midrule
    Binning $\binning$ & ``uniform" \\
    Maximum Bins, Univariate $\maxbins$ & 256 \\
    Maximum Bins, Bivariate $\maxbinstwod$ & $32\times32$ \\
    Learning Rate $\learningrate$ & 0.01 \\
    \bottomrule
    \end{tabular}
    \caption{Hyperparameters used to train the InterpretML \citep{nori2019}
             implementation of the EBM. All other model
             hyperparameters were set to the
             default values for InterpretML version 0.2.7.}
    \label{table:train_params}
\end{table}

\begin{table}
    \centering
    \begin{tabular}{c c c}
    \multicolumn{3}{c}{EBM Training Results} \\
    \toprule
    Metrics & $\ebmsfr$ & $\ebmms$\\
    \cmidrule(lr){2-2} \cmidrule(lr){3-3} \\[-12pt]
    $r^{2}$ & $0.898 \pm 0.0003$ & $0.882 \pm 0.0001$ \\
    $\zeta$ & $0.029 \pm 0.004$  & $0.008 \pm 0.0010$ \\[2pt]
     & $\log_{10}\sfr$ [$M_{\odot}yr^{-1}$] & $\log_{10}\mstar$ [$M_{\odot}$] \\
    \cmidrule(lr){2-2} \cmidrule(lr){3-3} \\[-12pt]
    MAE     & $0.144 \pm 0.0001$ & $0.189 \pm 0.0001$ \\

    \bottomrule
    \end{tabular}
    \caption{Training results for the EBM using $k$-fold cross
             validation. See Section \ref{sec:training} for more information
             on the training process. Reported are values for the variance
             metric $r^{2}$, the outlier fraction $\zeta$, and the mean absolute
             error (MAE). Uncertainties are computed from the
             variation among the $k$-fold trials.}
    \label{table:ebm_train_results}
\end{table}

\section{Results}
\label{sec:results}

After training the EBM model to reproduce the dependence of the
target quantities $\mstar$ and $\sfr$ on the parameters $\vtheta$,
the relationships between the target quantities and the parameters
can be analyzed. Below, we provide several analyses that
quantify how the target quantities relate to the parameters
and illustrate the performance of the EBM for our
astrophysical applications.

\subsection{Average Contribution}
\label{sec:avg_cont}

A key advantage of using EBM models over ``black box''
models (e.g., neural networks)
is their clear interpretability (see
\S \ref{sec:ebm}). The contribution of each
parameter $\theta_i$ to the model of the target quantity $y$
is provided by the
output functions $\fyi$ and $\fyij$.

Since these functions are vectors or two-dimensional
matrices with a number of elements equal to the number of bins $n_b$ in the
piece-wise function (see Table \ref{table:train_params}), a
summary scalar quantity for each feature function is helpful
for comparing their relative importance. We can
define the \emph{average contribution} $\cfave_y^i$ that
provides the average absolute value of $\fyi$ or $\fyij$,
with the average computed over the number of bins $n_b$
and weighted by the number of samples in each bin.
Mathematically, we can write
\begin{equation}
    \label{eqn:average_contribution}
    \cfave_y^i =  \frac{\sum_{j=0}^{n_b-1} |f(\theta_{i,j})|N_j}{\sum_{j=0}^{n_b-1}N_j}
\end{equation}
\noindent
where $f$ is the feature function being averaged ($\fyi$ or $\fyij$ from Equation
\ref{eqn:ebm}), $\theta_{i,j}$ is value of the parameter $\theta_i$
in the $j$th bin, and $N_j$
is the number of samples in bin $j$.
Intuitively, the average contribution $\cfave_y^i$
summarizes the importance of each parameter $\theta_i$
for determining the target quantities
when averaged over the samples in the final, merged dataset.

The
average contributions of each feature ($\fyi$) or combination of features
($\fyij$) are computed from the EBM. In each case, we rank order
the features by decreasing average contribution and
focus on the
seven features or feature combinations with the largest average contribution.
In each case the most important
feature has an average contribution more than
an order of magnitude larger than the seventh-ranked feature.

\subsubsection{EBM Model Targeting Star Formation Rate $\sfr$}
\label{sec:ebm_sfr}

\begin{figure}[h]
    \includegraphics[width=\linewidth]{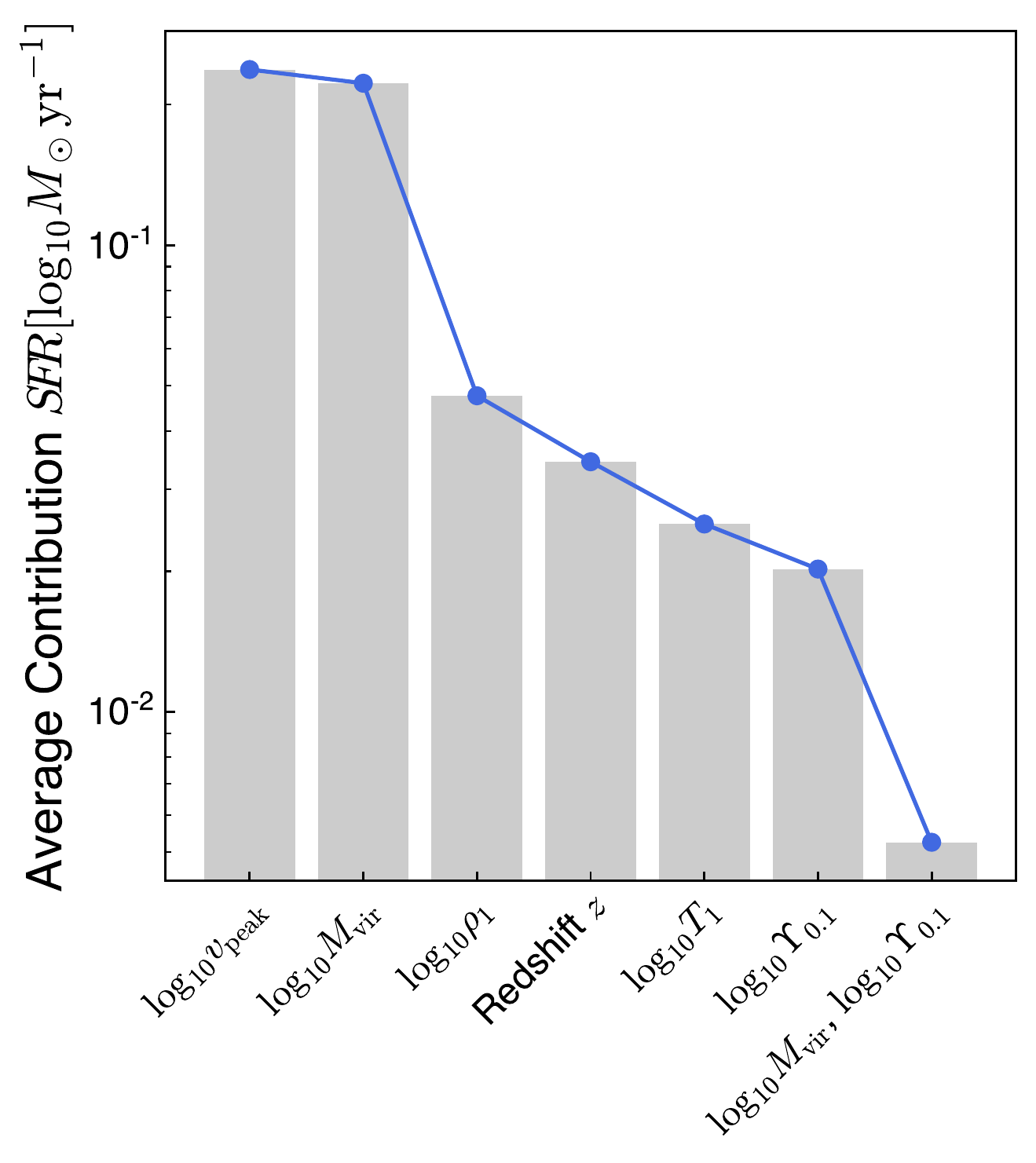}
    \caption{Top seven features with the highest average contribution in
            the EBM $\ebmsfr$ targeting the star formation rate $\sfr$. In order of
            decreasing importance, these features include peak circular
            velocity $\vpeak$, virial mass $\mvir$, environmental
            density $\rhoR$, redshift $z$, environmental temperature
            $\TR$, the mass ratio of nearby halos $\upR$,
            and the interaction between virial mass $\mvir$
            and $\upR$.
            Average contribution is
            calculated using the average of the absolute value of the
            feature functions weighted by the number of samples in each bin (see
            Equation \ref{eqn:average_contribution}).}
    \label{fig:sfr_ebm_avg_cont}
\end{figure}

Figure \ref{fig:sfr_ebm_avg_cont} shows the average
contribution of the top seven features
for the EBM model targeting star formation rate
$\log_{10} \sfr$.
In
decreasing order, the seven most important
features in determining $\sfr$ are
maximum peak circular velocity $\vpeak$,
virial mass $\mvir$, environmental
density $\rhoR$, redshift $z$,
environmental temperature $\TR$,
mass ratio of nearby halos $\upR$,
and the interaction between $\mvir$ and $\upR$.
The numerical values for the average contributions
are provided in Table \ref{table:sfr_ebm_avg_cont}.
The baseline value of $\sfr$ is $\beta_{\log_{10}\sfr} = -2.1151~[\log_{10}\Msun~\yr^{-1}]$, typical of halos with $\log_{10} \mvir\sim9$.
The average contribution of $\vpeak$ and $\mvir$ are quite similar,
providing $\Delta \log_{10} \sfr > 0.2$ on average, but their
interaction term is small with
$\cfave(\log_{10} \vpeak, \log_{10} \mvir)\ll0.01$.
Therefore peak circular velocity and virial mass provide
important contributions to
determining the star formation rate, and the
univariate dependence of the $\sfr$ on these properties
accounts for most of their contribution.
At the few-percent level, environmental density, redshift,
environmental gas temperature, and the presence of nearby massive
halos also contribute.

\begin{table}
    \centering
    \begin{tabular}{l c}
    \multicolumn{2}{c}{Average Contributions for the $\ebmms$ EBM} \\
    \toprule
    Feature & Value [$\log_{10}\Msun~\yr^{-1}$]\\
    \midrule
    $\beta_{\log_{10}\mstar}$                             & $-2.1151$ \\
    $\cfave(\log_{10} \vpeak)$                         & $0.2380$ \\
    $\cfave(\log_{10} \mvir)$                          & $0.2224$ \\
    $\cfave(\log_{10} \rhoR)$                        & $0.0475$ \\
    $\cfave(z)$                                        & $0.0343$ \\
    $\cfave(\log_{10} \TR)$                          & $0.0252$ \\
    $\cfave(\log_{10} \upR)$                    & $0.0202$ \\
    $\cfave(\log_{10} \mvir, \log_{10} \upR)$   & $0.0052$ \\
    \bottomrule
    \end{tabular}
    \caption{Summary of the EBM model trained to predict $\sfr$.
             The first entry, $\beta_{\log_{10}\sfr}$, is the
             baseline value learned model (see Section \ref{sec:ebm}). The next seven
             entries are the average contributions of the most important feature functions listed
             in descending order (see Equation \ref{eqn:average_contribution}).
             }
    \label{table:sfr_ebm_avg_cont}
\end{table}

\begin{figure*}
    \includegraphics[width=\textwidth]{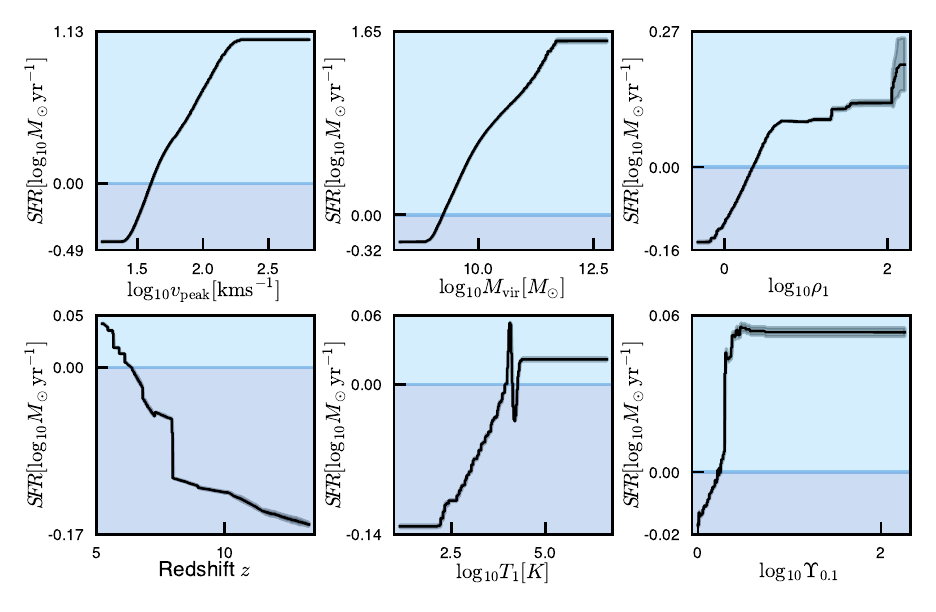}
    \caption{Learned univariate feature functions $\fyi$
             for the EBM $\ebmsfr$ trained to predict the star formation rate
             $\sfr$.
             Shown (left to right)
             are the feature functions for peak circular
             velocity
             $\vpeak$, virial mass $\mvir$,
             environmental density $\rhoR$,
             redshift $z$, environmental temperature $\TR$, and nearby halo mass ratio
             $\upR$.
             Light blue areas indicate regions where $\fyi>0$ and
             dark blue areas indicate regions where $\fyi<0$. The shaded areas
             show the variation in $\fyi$ between the $k$-fold iterations.}
    \label{fig:sfr_ebm_univariate}
\end{figure*}

The feature functions $\fyi$ for each feature are plotted in Figure
\ref{fig:sfr_ebm_univariate}. The functions indicate that there are
positive
correlations between the star formation rate
$\log_{10} \sfr$ and either the peak circular velocity
$\vpeak$,
virial mass $\mvir$, or environmental
density $\rhoR$. The star formation
rate increases with increasing environmental temperature $\log_{10}\TR$,
but near $\TR\approx10^{4}$K the univariate function shows an enhancement
just as hydrogen becomes mostly neutral and a deficit near the temperature
at which it becomes ionized.
Star formation rate increases with decreasing redshift over the range $z\sim5-15$,
becoming more efficient after reionization.

\begin{figure*}[h]
    \includegraphics[width=\linewidth]{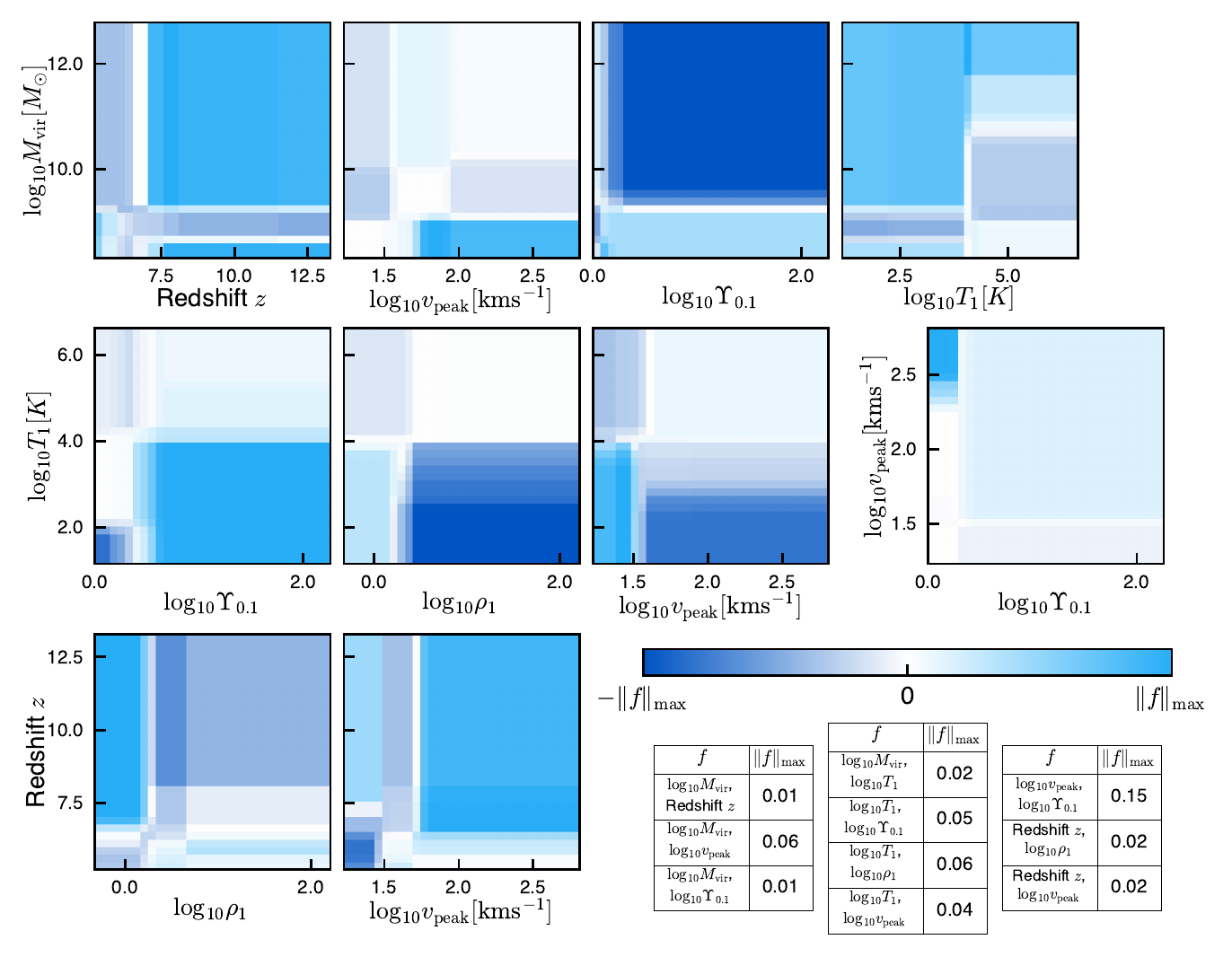}
    \caption{Most important learned interaction functions $\fyij$
             for the EBM model $\ebmsfr$ targeting the star formation
             rate $\sfr$, as a function of their parameter pairs. Each panel
    shows the contribution of the bivariate interaction terms, normalized
    such that the color map ranges between plus or minus the maximum of
    the norm of each function $||f||_\mathrm{max}$.  Light blue areas
             indicate regions of joint parameter space where the
             feature interactions contribute positively to the star formation
             rate, while dark blue areas indicate regions with negative
             contributions.
    The table lists
    $||f||_\mathrm{max}$ for the interaction functions, each with
    units $\log_{10}\Msun~\yr^{-1}$. In absolute terms, the largest
    interaction occurs for halos with large peak circular velocity $\vpeak$
    and no large neighboring halos ($\upR\approx0$). The other interaction
    functions are relatively weak, and contribute changes to $\log\sfr \lesssim0.05$.
    }
    \label{fig:sfr_ebm_interaction}
\end{figure*}

The interaction functions $\fyij$ learned by the
EBM $\ebmsfr$ targeting the star formation rate $\sfr$ are plotted as
``heat maps'' in Figure
\ref{fig:sfr_ebm_interaction}. Most interaction functions do not contribute
significantly to the star formation rate, and change the star formation
rate by $\Delta \log_{10} \sfr \lesssim 0.05$. However, halos with low
neighboring halo mass ratios $\upR$ and large peak circular velocity
$\vpeak$ have their star formation rate enhanced by $\Delta \log_{10} \sfr \approx 0.15$.
Rephrased, locally dominant halos with large peak circular velocity
show enhanced star formation. Such enhancements likely owe to recent
merger activity.

\begin{figure}[h]
    \includegraphics[width=\columnwidth]{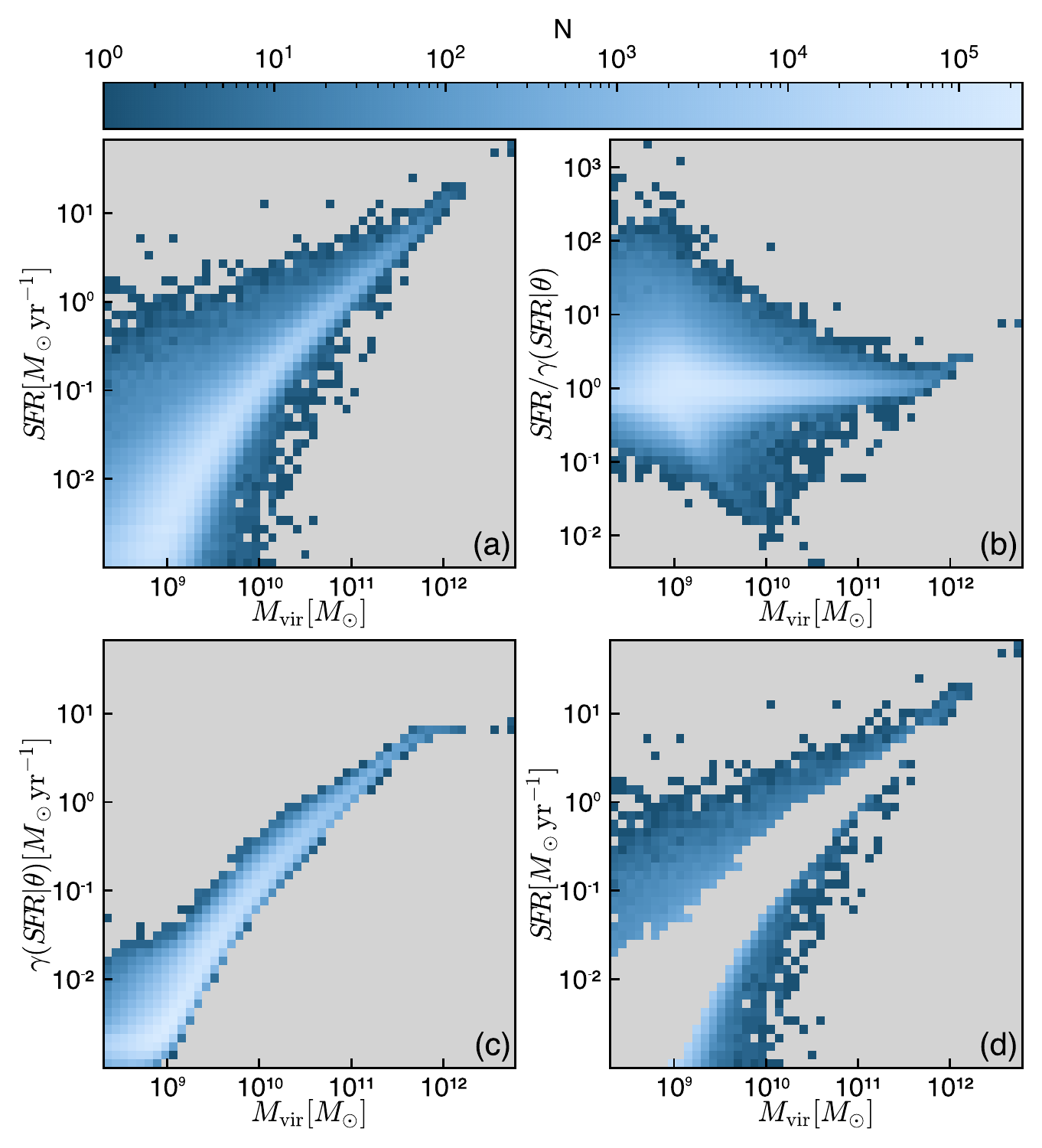}
    \caption{Summary of the EBM model $\ebmsfr$ targeting star formation rate ($\sfr$)
             as a function of virial mass. The upper left panel shows the two-dimensional
             distribution of $\sfr$ with $\mvir$ for galaxies in the CROC simulations, with the color
             scale showing the number of simulated galaxies at each $[\sfr,\mvir]$
             location. The lower left panel shows the EBM model results for the
             distribution of $\sfr$ with $\mvir$, where the $\sfr$ is computed
             from the EBM using the parameters $\vtheta = [\mvir, \vpeak, z, \rhoR, \TR, \upR]$.
             The upper right panel shows the residuals between the simulated CROC
             galaxy $\sfr$s and the EBM model results. The lower right panel shows
             the simulated CROC galaxy $\sfr$s that lie outside the EBM model predictions.
             These outliers represent $\lesssim3\%$ of simulated CROC galaxies.}
    \label{fig:sfr_ebm_model_summary}
\end{figure}

While Equation \ref{eqn:ebm} represents a complex, multidimensional
manifold that provides
the $\sfr$ as a function of the parameters $\vtheta$, the
distributions of simulated and predicted $\sfr$
as a function of a single parameter provide a graphical summary
of the EBM model performance.
Figure \ref{fig:sfr_ebm_model_summary} shows the simulated and
predicted $\sfr$ as a function of virial mass $\log_{10} \mvir$,
and we will refer to this figure as the \emph{model summary}.
Shown in this model summary
are the distributions of $\sfr$ in the CROC simulated galaxy
catalogs with virial mass
and the $\sfr$ predicted by the EBM model $\ebmsfr$
using the parameters $\vtheta$
measured for each simulated galaxy. The EBM model captures roughly
97\% of the simulated distribution of $\sfr$ with virial mass. The
EBM model is highly predictive of the simulated connection between
$\sfr$ and the intrinsic and extrinsic properties $\vtheta$.

Given the combined complexity of the average contribution measures,
univariate feature functions, and bivariate interaction functions,
in what follows we will show the summary figure for other EBM models
in the main text. For completeness, the average contribution,
feature function, and interaction function figures for each model
will be presented in the Appendices.

\subsubsection{EBM Model Targeting Stellar Mass $\mstar$}
\label{sec:ebm_mstar}

An EBM model $\ebmms$
targeting stellar mass $\mstar$ using the
properties $\vtheta$ can be constructed through simple
retraining. Using the simulated galaxy catalogs from CROC,
we retrain the EBM to model $\mstar$ against $\vtheta$.
We report the average contribution, univariate feature
functions, and bivariate interaction functions for $\ebmms$
in Appendix \ref{appendix:ebm_mstar}.
For reference,
the baseline value of $\mstar$ is $\beta_{\log_{10}\mstar} = 5.9629~[\log_{10}\Msun~\yr^{-1}]$ (see Table \ref{table:mstar_ebm_avg_cont} in the Appendix), typical of halos with $\log_{10} \mvir/\Msun\sim9$.

\begin{figure}[h]
    \includegraphics[width=\columnwidth]{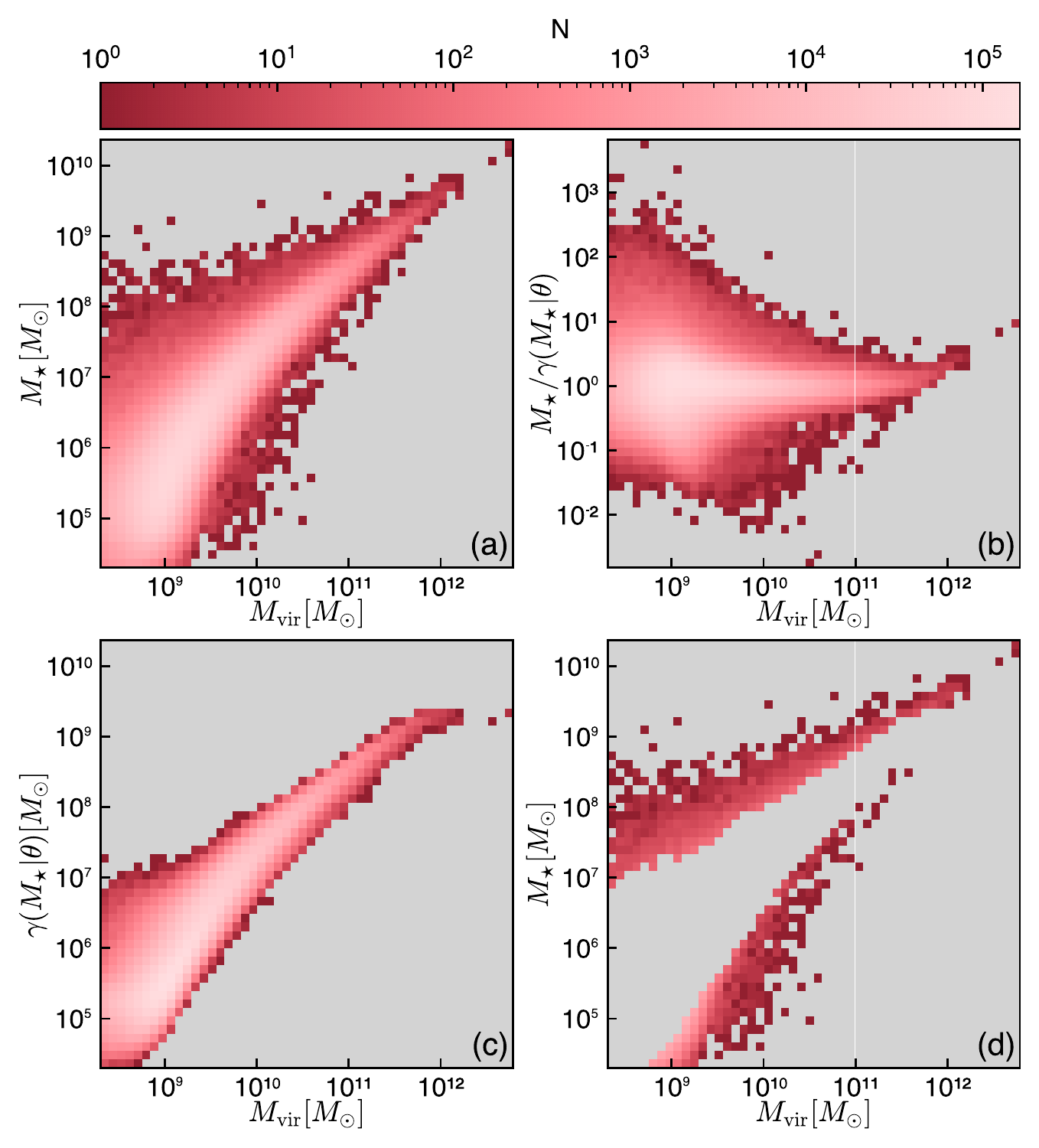}
    \caption{Summary of the EBM model $\ebmms$ targeting stellar mass $\mstar$
    as a function of virial mass.
    The upper
    left panel shows the distribution of $\mstar$
    with virial mass $\mvir$ in the
    CROC simulated galaxy catalogs, with the coloration indicating the number of
    galaxies at each $[\mstar, \mvir]$ location.
    The lower left panel shows the EBM model
    prediction of the stellar mass distribution with virial mass given in the
    input parameters $\vtheta = [\mvir, z, \vpeak, \rhoR, \TR, \upR]$. The
    upper right panel shows the residuals between the simulated and predicted
    $\mstar$ vs. $\mvir$ distribution, and the lower right panel shows the
    outliers in the simulated distribution not captured by the EBM model
    $\ebmms$. The fraction of outliers is $\lesssim1\%$.}
    \label{fig:mstar_ebm_model_summary}
\end{figure}

Figure \ref{fig:mstar_ebm_model_summary} shows the model
summary for the EBM model $\ebmms$. The EBM
model provides an excellent representation of the
distribution of stellar masses for the CROC
simulated galaxy catalog. As the lower
right panel of Figure \ref{fig:mstar_ebm_model_summary}
indicates, the $\ebmms$ model results in few
outliers for the CROC simulated galaxies and has an
outlier fraction of $\lesssim1\%$. Given the galaxy
properties $\vtheta = [\mvir, z, \vpeak, \rhoR, \TR, \upR]$,
the distribution of stellar masses for CROC simulated galaxies can
be recovered to 99\% accuracy.

\section{Composite EBMs for Restricted Parameter Sets}
\label{sec:cebm}

The EBM models $\ebmsfr$ and $\ebmms$
presented in \S \ref{sec:ebm_sfr} and \S \ref{sec:ebm_mstar}
are constructed using the parameter set
$\vtheta = [\mvir, z, \vpeak, \rhoR, \TR, \upR]$. Our
results show that the full distribution of $\sfr$ and stellar
mass in the simulated CROC galaxy catalogs can be
recovered accurately with only $\approx1-3\%$ outliers.
These EBM models can therefore be applied to cosmological simulations
using the parameters $\vtheta$ measured from simulated
galaxy catalogs to recover the distribution of $\sfr$
and stellar mass computed by CROC.

The parameters $\vtheta$ include the peak circular
velocity $\vpeak$, which requires both time-dependent
tracking of formation histories for individual
galaxies and high spatial resolution to capture the
peak of the rotation curve for each object. As a result,
as expressed above the models $\ebmsfr$ and
$\ebmms$ cannot be applied directly to
cosmological simulations with low spatial resolution
or without merger trees to capture formation histories.

Instead of fitting EBM models using the full parameter
set $\vtheta$, consider the construction of an EBM
model using the restricted parameter set
$\vtheta' = [\mvir, z, \rhoR, \TR, \upR]$ that
does not include $\vpeak$. The parameters $\vtheta'$
can all be measured directly in cosmological
simulations with sufficient resolution to capture
individual galaxy-mass halos without the need to
track merger trees. The EBM models $\ebmsfrp$
and $\ebmmsp$ using the restricted
parameter set $\vtheta'$ perform substantially less
well than the models $\ebmsfr$
and $\ebmms$ trained on the full parameter
set $\vtheta$ that includes $\vpeak$. With the
restricted parameter set $\vtheta'$, the EBM model
shows $7.6\%$ outliers when targeting $\sfr$ and
$2.8\%$ when targeting $\mstar$. Comparing with
the outlier fractions reported in
Table \ref{table:ebm_train_results} for the full
parameter set including $\vpeak$, the EBM model
trained on the restricted dataset has degraded
its performance by a factor of $\sim2-3$.

To ameliorate the poorer performance of the EBM models
trained on restricted parameter sets, we use a
\emph{Composite EBM} (CEBM) model.
Given a target quantity $y$ and a parameter set $\vtheta'$,
we fit a \emph{base} EBM $\basey$ in the same manner as
fitting the EBMs $\ebmsfr$ or $\ebmms$. We construct
a dataset from the galaxies whose $y$ values lie outside
the predictions from $\basey$, and then fit an \emph{outlier}
EBM $\outy$ to these discrepant samples. We then weight
the base and outlier EBMs to construct the CEBM model $\cebmy$
using a \emph{classifier} EBM $\classy$. Instead of fitting
the change in star formation rate or stellar mass at a given sample
in $\theta'$, the classifier EBM fits the log odds that a
given sample in $\theta'$ is an outlier. We then define
$\classy$ to be the sigmoid of these log odds, such that
$\classy\in [0,1]$. The CEBM can then be written
as
\begin{equation}
\label{eq:cebm}
\cebmy = [1-\classy]\basey + \classy\outy.
\end{equation}
We describe the CEBM approach in more detail in Appendix \ref{appendix:cebm},
and provide information on the CEBMs $\cebmsfr$ and $\cebmms$ in
Appendices \ref{appendix:cebm_sfr} and \ref{appendix:cebm_mstar}.

Table \ref{table:cebm_train_results} lists the evaluation metrics
for the training of CEBM models targeting $\sfr$ and stellar
mass without using $\vpeak$. The outlier fraction
has improved to $\approx5\%$ for
CEBM model $\cebmsfr$ and to $\lesssim2\%$ for
$\cebmms$. The average parameter contributions
and baseline value $\beta_{\log_{10} \sfr}$ from $\cebmsfr$
are provided in Table \ref{table:sfr_composite_ebm_overall} and
for the CEBM targeting stellar mass in Table \ref{table:mstar_composite_ebm_overall}. The univariate feature functions
and bivariate interaction functions for the CEBM models
$\cebmsfr$ and $\cebmms$ are provided in
Appendices \ref{appendix:cebm_sfr} and \ref{appendix:cebm_mstar}.

\begin{figure}[h]
    \includegraphics[width=\columnwidth]{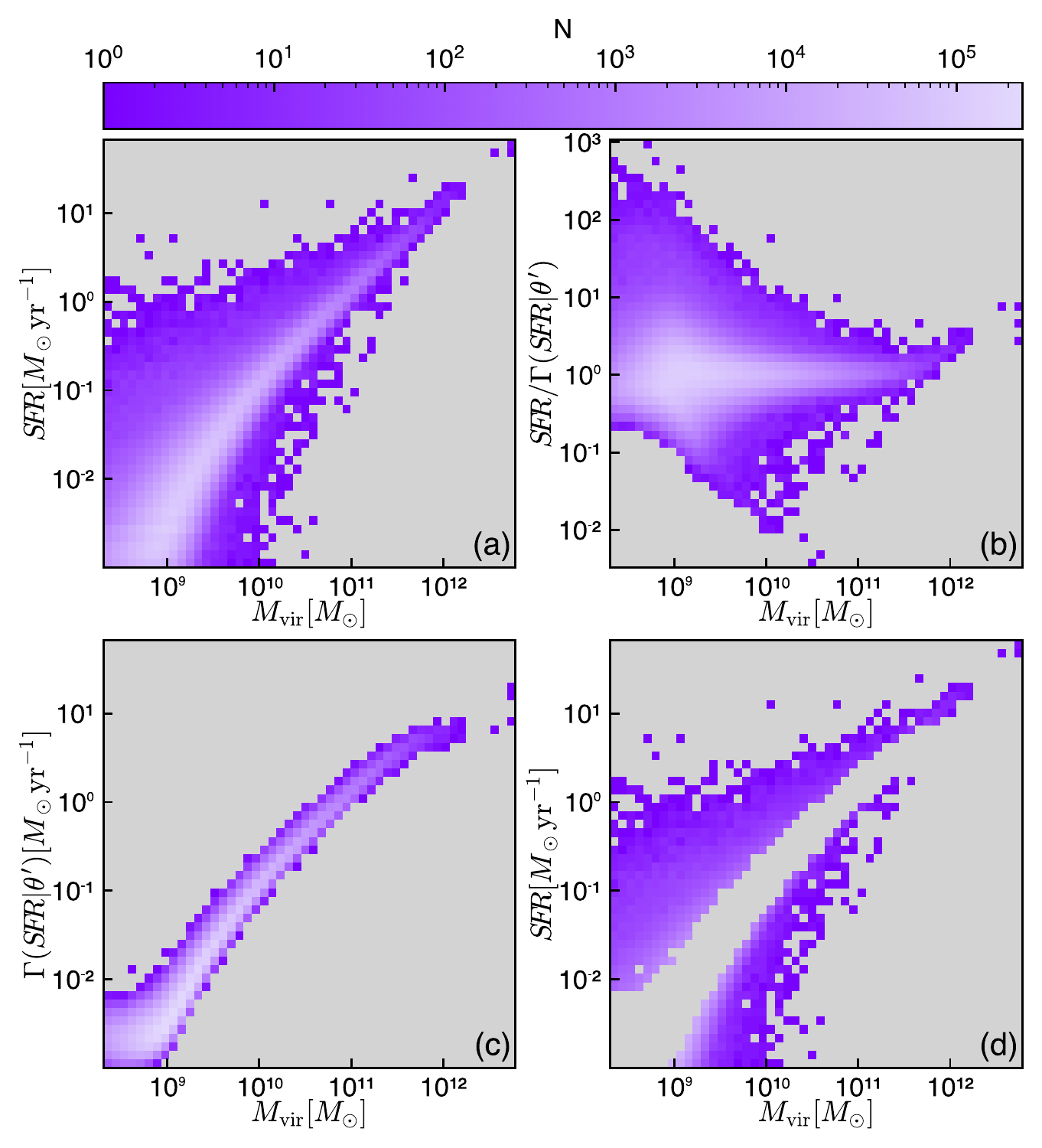}
    \caption{Summary of the CEBM model $\cebmsfr$ targeting star formation rate ($\sfr$)
             as a function of virial mass.
             The upper left panel shows the two-dimensional
             distribution of $\sfr$ with $\mvir$ for galaxies in the CROC simulations, with the color
             scale showing the number of simulated galaxies at each $[\sfr, \mvir]$
             location. The lower left panel shows the CEBM model results for the
             distribution of $\sfr$ with $\mvir$, where the $\sfr$ is computed
             from the CEBM using the parameters $\vtheta' = [\mvir, z, \rhoR, \TR, \upR]$.
             The upper right panel shows the residuals between the simulated CROC
             galaxy $\sfr$s and the CEBM model results. The lower right panel shows
             the simulated CROC galaxy $\sfr$s that lie outside the CEBM model predictions.
             These outliers represent $\approx5\%$ of simulated CROC galaxies.}
    \label{fig:sfr_cebm_model_summary}
\end{figure}

\begin{figure}[h]
    \includegraphics[width=\columnwidth]{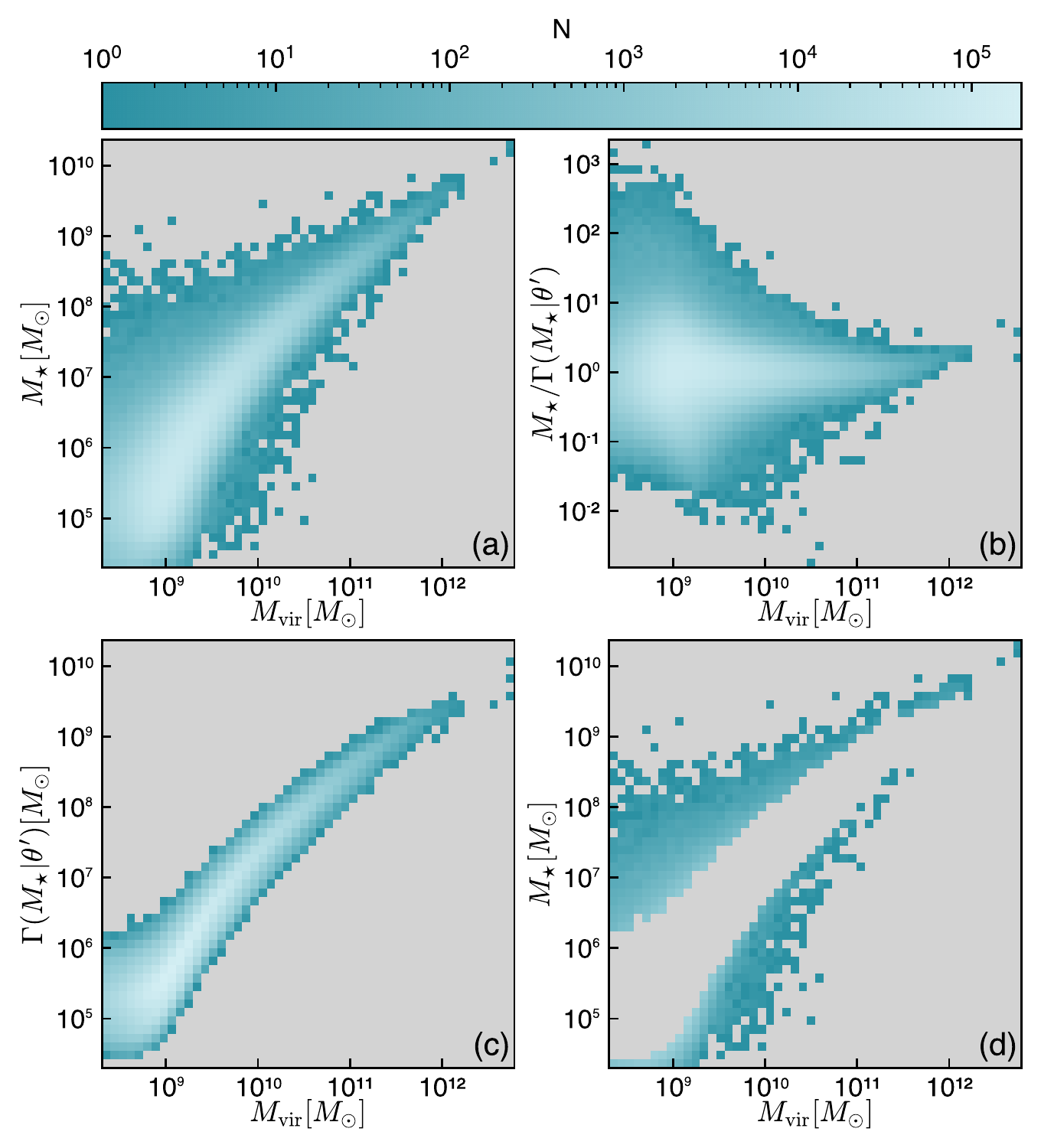}
    \caption{Summary of the CEBM model $\cebmms$ targeting stellar mass $\mstar$
             as a function of virial mass.
    The upper
    left panel shows the distribution of $\mstar$ with virial mass $\mvir$ in the
    CROC simulated galaxy catalogs, with the coloration indicating the number of
    galaxies at each $[\mstar, \mvir]$ location.
    The lower left panel shows the CEBM model
    prediction of the stellar mass distribution with virial mass given in the
    input parameters $\vtheta' = [\mvir, z, \rhoR, \TR, \upR]$. The
    upper right panel shows the residuals between the simulated and predicted
    $\mstar$ vs. $\mvir$ distribution, and the lower right panel shows the
    outliers in the simulated distribution not captured by the CEBM model
    $\cebmms$. The fraction of outliers is $\lesssim2\%$.}
    \label{fig:mstar_cebm_model_summary}
\end{figure}

Figure \ref{fig:sfr_cebm_model_summary} shows the model summary for the
CEBM targeting $\sfr$, and Figure \ref{fig:mstar_cebm_model_summary} shows the model summary for the CEBM targeting stellar mass. As both models
demonstrate, the CEBM model accurately recovers the distribution of
star formation rate and stellar mass in the CROC simulated galaxy
sample. Between the models, the outlier fraction is only $\approx2-5\%$
despite using the restricted set of parameters $\vtheta'$ that does
not include $\vpeak$ or any time-dependent tracking of individual
systems.

\begin{table}
    \centering
    \begin{tabular}{c c c}
    \multicolumn{3}{c}{Composite EBM Training Results} \\
    \toprule
    Metrics & $\ebmsfr$ & $\ebmms$\\
    \cmidrule(lr){2-2} \cmidrule(lr){3-3} \\[-12pt]
    $r^{2}$ & $0.868 \pm 0.0002$ & $0.830 \pm 0.0003$ \\
    $\zeta$ & $0.052 \pm 0.0053$  & $0.018 \pm 0.0031$ \\[2pt]
     & $\log_{10}\sfr$ [$M_{\odot}yr^{-1}$] & $\log_{10}\mstar$ [$M_{\odot}$] \\
    \cmidrule(lr){2-2} \cmidrule(lr){3-3} \\[-12pt]
    MAE     & $0.165 \pm 0.0001$ & $0.233 \pm 0.0002$ \\
    \bottomrule
    \end{tabular}
    \caption{Training results for CEBM models for $\sfr$ and $\mstar$
    using $k$-fold cross
             validation. See Section \ref{sec:training} for more information
             on the training process. Reported are values for the variance
             metric $r^{2}$, the outlier fraction $\zeta$, and the mean absolute
             error (MAE). Uncertainties are computed from the
             variation among the $k$-fold trials.}
    \label{table:cebm_train_results}
\end{table}

\begin{table}
    \centering
    \begin{tabular}{l c}
    \multicolumn{2}{c}{Overview of CEBM $\cebmsfr$} \\
    \toprule
    Feature & Value [$\log_{10} \Msun~\yr^{-1}$]\\
    \midrule
    $\beta_{\log_{10}\sfr}$                             & $-1.7466$ \\
    $\cfeat(\log_{10} \mvir)$                          & $0.4327$ \\
    $\cfeat(\log_{10} \rhoR)$                        & $0.0625$ \\
    $\cfeat(\log_{10} \TR)$                          & $0.0327$ \\
    $\cfeat(\log_{10} \upR)$                    & $0.0215$ \\
    $\cfeat(z)$                                        & $0.0190$ \\
    $\cfeat(z, \log_{10} \rhoR)$                     & $0.0077$ \\
    $\cfeat(\log_{10} \mvir, \log_{10} \upR)$   & $0.0056$ \\
    \bottomrule
    \end{tabular}
    \caption{Average contribution to the CEBM model $\cebmsfr$
             trained to predict $\sfr$ from
             the parameter set $\vtheta'$.
             The first entry, $\beta_{\log_{10}\sfr}$, is
             the learned baseline of the model. The next seven entries are the
             feature functions with the highest average contribution listed in
             descending order. The average contribution is calculated using the
             average of the absolute value of the base EBM function
             values weighted by the number of samples in each bin and the output
             of the classification EBM for each sample (see
             Appendix
             \ref{appendix:cebm_average_contribution} for more details).}
    \label{table:sfr_composite_ebm_overall}
\end{table}

\section{Discussion}
\label{sec:discussion}

Explainable Boosting Machine (EBM) models provide a method to
statistically
infer relationships present in high-dimensional data.
Given their statistical nature,
EBM models remain ignorant of the physics that generate the
connection between star formation rate, stellar mass,
and the properties of dark matter halos that host galaxies.
Nonetheless, given the results of detailed physical modeling
in the form of simulated galaxy catalogs from cosmological simulations,
the EBM correctly identifies halo mass and maximum peak circular velocity
as the most important halo properties for determining
$\sfr$ and $\mstar$ (e.g., Figure \ref{fig:sfr_ebm_avg_cont}).
The EBM correctly infers that $\sfr$ and $\mstar$ increase
with increasing halo mass or $\vpeak$, and the EBM
univariate feature functions correctly identify the
gas temperature at which star formation efficiency changes.
To the extent that the physical connection between
galaxy and halo properties are recorded in
statistical relationships, the EBM models
effectively recover some fraction of those relations.

EBM models also provide a means to implement a
``sub-grid'' prescription for galaxy formation
based on the properties of halos and their environments.
The EBM models $\ebmsfr$ and $\ebmms$
capture better than 97\% of the $\sfr$ and $\mstar$ distributions
measured for simulated galaxies in the CROC simulations.
The stellar masses and star formation rates of galaxies in
CROC could be accurately recovered by using only the
halo and environmental parameters in
$\vtheta = [\mvir, z, \vpeak, \rhoR, \TR, \upR]$.

Using the CEBM model trained on the restricted
parameter set $\vtheta' = [\mvir, z, \rhoR, \TR, \upR]$,
$\approx95-98\%$ of the distribution
of $\sfr$ and $\mstar$ of the CROC galaxies is recovered.
One advantage of this parameter set is that the
spatial resolution in the simulations required
to compute them is less demanding than for $\vpeak$. A simulation
with coarser resolution than CROC, such that the
details of the star formation and feedback processes
cannot be resolved, may still leverage the
CEBM models $\cebmsfr$ and $\cebmms$
to model the star formation rate and stellar masses
in dark matter halos. Further, the quantities $\vtheta'$
used
to train the CEBM models are measured at distinct
redshifts such that no merger trees are required
to recover accurately the CROC $\sfr$ and $\mstar$
distributions from halo and environmental properties.
We note that
for both the EBM and CEBM models the outlier fractions
not well captured by the model are
roughtly percent-level or less in the $\sfr$ or
$\mstar$ distributions, and we expect that
corresponding inaccuracies induced in, e.g., the ionizing photon
budget or topology of reionization will be minimal.

By editing the dataset and retraining, the impact of
environment on the performance of the EBM models
can be estimated. Relative to $\ebmsfr$ and $\ebmms$
that use the full dataset $\vtheta$ including all
environmental parameters, EBM models trained only
on maximum peak circular velocity $\vpeak$,
halo virial mass $\mvir$, and redshift $z$ have
an outlier fraction increased by only $\sim1\%$
when modeling $\mstar$ and $\sim10\%$ when
modeling $\sfr$. Further, removing $\vpeak$
and training only on $[\mvir,z]$ substantially
degrades the model performance, and the
outlier fractions increase to $\sim20\%$ when
modeling $\mstar$ and $\sim40\%$ when modeling
$\sfr$. The importance of including $\vpeak$
in the training dataset is much larger than
the importance of accounting for
the environmental measures selected in this
analysis.

The EBM models enable an approximate translation of the
galaxy formation model from one simulation to another.
Provided the parameter sets $\vtheta$ or $\vtheta'$ can
be measured in both simulations, an EBM can recover
the connection between $\sfr$, stellar masses, halo
properties, and environment from the training
simulation and then be used to instill those relations
in a different simulation.
Since the $\vtheta'$ parameter
set does not require very high spatial resolution to
capture, the net results for $\sfr$ and stellar mass
from a high resolution simulation accurately tracking
detailed baryonic physics can be translated into
a simulation with resolution insufficient to
capture those physics directly. In future work, we
plan to transfer the CROC baryonic galaxy formation
model into \emph{Cholla} cosmological simulations
\citep[e.g.,][]{villasenor2021a,villasenor2022a}
via the EBM models presented here. Such a transferred
model could be used to build models of feedback from
galaxy formation on resolved scales that incorporate
the regulatory effects of feedback on small-scale
star formation.

Lastly, the ability of the EBM models to recover the
$\sfr$ and $\mstar$ distributions using
only halo and environmental properties
allows for the rapid replacement of galaxy formation
models based on EBMs. Models can be trained on
the simulated galaxy catalogs from a variety of
expensive, high-resolution training simulations including a
wide range of physics. These EBM models can then
be used interchangeably as effective galaxy formation
models in the target simulations, and can also be modified posteriori to allow a broad parameter search or
correct the inaccuracies of the training simulation.
Such an
approach could reduce the sensitivity of conclusions
about, e.g., the reionization process on the
detailed $\sfr$ and $\mstar$ distributions as multiple
EBM models for these properties could be trained
and implemented in the target simulations.

\section{Summary}
\label{sec:summary}

A complex interplay of physical processes gives rise to the distribution
of star formation rates ($\sfr$s) and stellar masses $\mstar$
of galaxies over cosmic
time. Cosmological simulations provide powerful methods for modeling these
physical processes, but the connection between $\sfr$, $\mstar$, and other
galaxy properties can be obfuscated by complexity. Leveraging machine
learning techniques, we use a variation of the Generalized
Additive Model \citep{Hasti1986}
called Explainable Boosting Machines \citep[EBM][]{nori2019} to infer the
dependence of $\sfr$ and $\mstar$ in the Cosmic Reionization on Computers
(CROC) simulations \citep{gnedin2014a} on dark matter halo properties
including virial mass $\mvir$,
peak maximum circular velocity $\vpeak$, redshift,
environmental density, environmental gas temperature, and the
mass of neighboring halos. Our findings include:

\begin{itemize}
  \item $\sfr$ and $\mstar$ primarily depend on $\mvir$ and $\vpeak$, followed
  by redshift, environmental density, and environmental gas temperature.
  \item When including $\mvir$ and $\vpeak$ in the parameter set used to
  train the EBM, the model recovers better than 97\% of the distribution
  of $\mstar$ or $\sfr$ with virial mass $\mvir$ in the CROC simulations.
  \item If the model fit excludes $\vpeak$, the fraction of outliers
  in the CROC data relative to the predicted model distribution increases
  to 7.6\% for $\sfr$ and 2.8\% for $\mstar$.
  \item To ameliorate the degradation of the model performance when excluding $\vpeak$,
  we define a composite EBM model comprised of a weighted sum of the
  base EBM
  model fit to main trend of $\sfr$ and $\mstar$ with the halo properties and
  a second EBM model to fit the outliers not represented in the base EBM.
  The weighting coefficients are themselves determined by an EBM model fit.
  \item The composite EBM model improves the performance to $\approx95-98\%$
  accuracy in the distribution of $\sfr$ or $\mstar$ with virial mass, even
  when excluding $\vpeak$ measurements from the training dataset.
\end{itemize}

The EBM models quantify the relative importance of halo properties like
virial mass and maximum peak circular velocity for determining
the stellar mass and star formation rate of the galaxy it hosts. Through
these models, the physics of baryonic galaxy formation can be connected
to the properties of dark matter halos and enable galaxy formation to
be implemented as a ``sub-grid'' prescription in dark matter-only
simulations or hydrodynamical simulations that do not resolve the
small scale details of star formation and feedback.

\begin{acknowledgments}

This work was supported by the NASA Theoretical and Computational Astrophysics
Network (TCAN) grant 80NSSC21K0271. The authors acknowledge use of the lux
supercomputer at UC Santa Cruz, funded by NSF MRI grant AST 1828315. This
manuscript has been co-authored by Fermi Research Alliance, LLC under Contract
No. DE-AC02-07CH11359 with the U.S. Department of Energy, Office of Science,
Office of High Energy Physics. CROC project relied on resources of the Argonne
Leadership Computing Facility, which is a DOE Office of Science User Facility
supported under Contract DE-AC02-06CH11357. An award of computer time was
provided by the Innovative and Novel Computational Impact on Theory and
Experiment (INCITE) program. CROC project is also part of the Blue Waters
sustained-petascale computing project, which is supported by the National
Science Foundation (awards OCI-0725070 and ACI-1238993) and the state of
Illinois. Blue Waters is a joint effort of the University of Illinois at
Urbana-Champaign and its National Center for Supercomputing Applications. This
research used resources of the Oak Ridge Leadership Computing Facility, which is
a DOE Office of Science User Facility supported under Contract
DE-AC05-00OR22725. We have used resources from DOE INCITE award AST 175.

\end{acknowledgments}

\software{
    Python \citep{rossum1995},
    NumPy \citep{vanderwal2011},
    scikit-learn \citep{pedregosa2011},
    matplotlib \citep{hunter2007},
    InterpretML \citep{nori2019},
}

\bibliography{ref}

\appendix
\section{Detailed Results for the $\mstar$ EBM}
\label{appendix:ebm_mstar}

While the performance of the
EBM model $\ebmms$ targeting
$\mstar$
is summarized in Figure \ref{fig:mstar_ebm_model_summary},
a more detailed view of the model is provided by
the average contributions provided by each parameter,
the univariate feature functions dependent on the
parameters, and the bivariate interaction functions.
These results of the model are presented below.

\subsection{Average Contribution}

Figure \ref{fig:mstar_ebm_avg_cont} shows the average contribution
of the seven most important features and interactions
in the EBM model $\ebmms$. In order
of decreasing importance, these features include peak circular
velocity, virial mass, redshift, environmental density, environmental
temperature, the mass ratio of nearby halos, and the interaction
between redshift and peak circular velocity. Peak circular velocity
is about 50\% more important than virial mass, which in turn is
roughly a factor of two more important than redshift. The other
features and interactions contribute to
stellar mass at the $\lesssim 0.1$ dex level.
For reference the numerical values for the average contributions
are provided in Table \ref{table:mstar_ebm_avg_cont}.

\begin{figure}[h]
    \centering
    \includegraphics[width=0.5\textwidth]{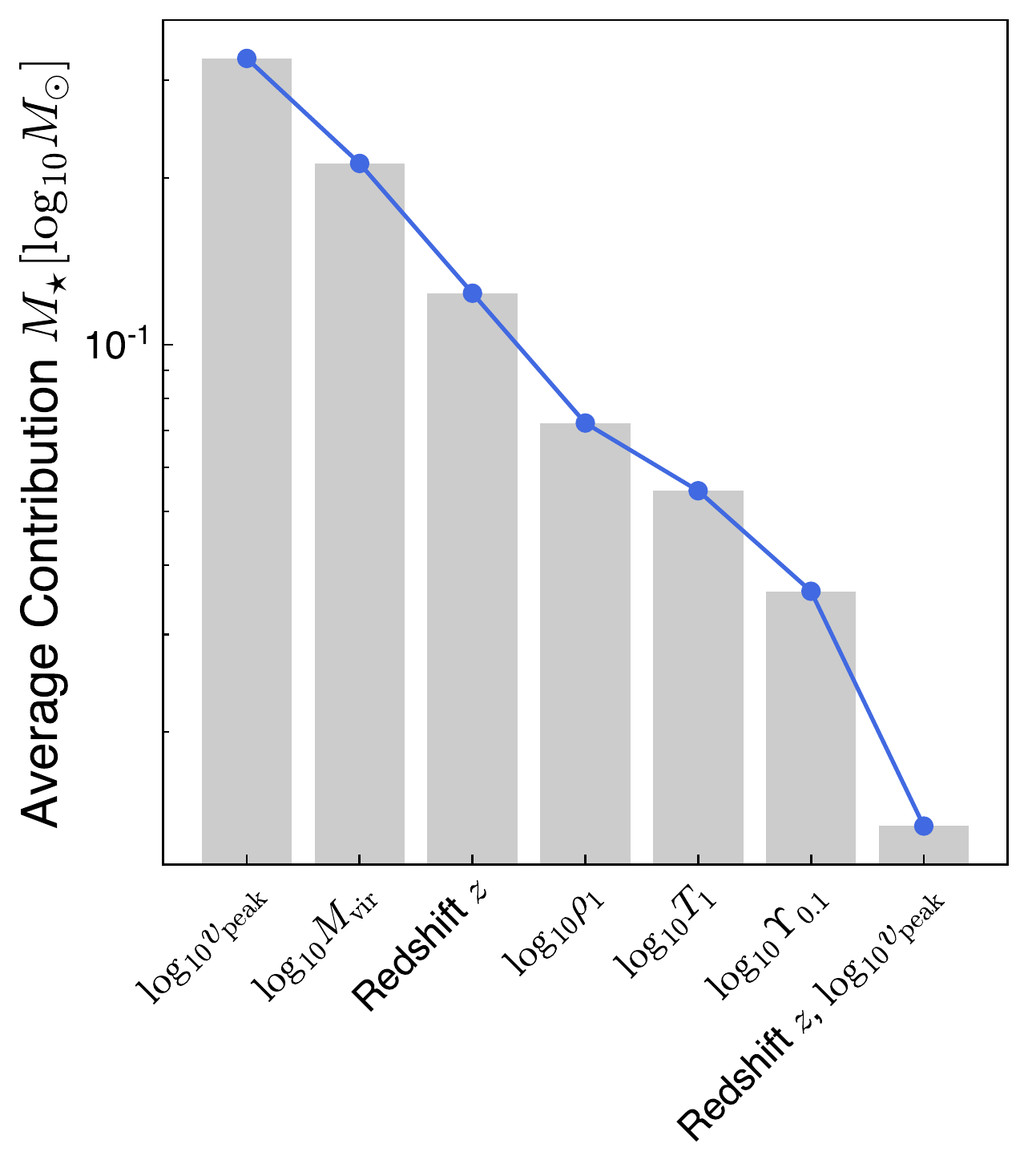}
    \caption{Features with the highest average contribution for
             the EBM $\ebmms$ trained to predict $\mstar$. Average contribution is
             calculated using the average of the absolute value of the learned
             functions weighted by the number of samples in each bin (see
             Equation \ref{eqn:average_contribution}). The features with
             the largest contributon are $\vpeak$ and $\mvir$, followed by
             redshift $z$, environmental density $\rhoR$, environmental
             temperature $\TR$, and mass ratio of nearby halos $\upR$. The
             interaction with the largest average contribution involves [$z$,$\vpeak$].}
    \label{fig:mstar_ebm_avg_cont}
\end{figure}

\begin{table}
    \centering
    \begin{tabular}{l c}
    \multicolumn{2}{c}{Average Contributions for the $\ebmms$ EBM} \\
    \toprule
    Feature & Value $[\log_{10} \Msun]$\\
    \midrule
    $\beta_{\log_{10}\mstar}$                           & $5.9629$ \\
    $\cfave(\log_{10} \vpeak)$                         & $0.3284$ \\
    $\cfave(\log_{10} \mvir)$                          & $0.2123$ \\
    $\cfave(z)$                                        & $0.1238$ \\
    $\cfave(\log_{10} \rhoR)$                        & $0.0722$ \\
    $\cfave(\log_{10} \TR)$                          & $0.0545$ \\
    $\cfave(\log_{10} \upR)$                    & $0.0359$ \\
    $\cfave(z, \log_{10} \vpeak)$                      & $0.0135$ \\
    \bottomrule
    \end{tabular}
    \caption{Summary of the EBM model $\ebmms$ trained to predict $\mstar$ as a
             function of the full parameter set $\vtheta$.
             The first entry, $\beta_{\log_{10}\mstar}$, is
             the learned baseline value of the model (see Section \ref{sec:ebm}).
             The next seven entries are the feature functions with the highest
             average contribution in descending order. Average contribution is
             calculated using the average of the absolute value of the feature
             functions weighted by the number of samples in each bin (see
             Equation \ref{eqn:average_contribution}).}
    \label{table:mstar_ebm_avg_cont}
\end{table}

\subsection{Feature Functions}

\begin{figure*}
    \includegraphics[width=\textwidth]{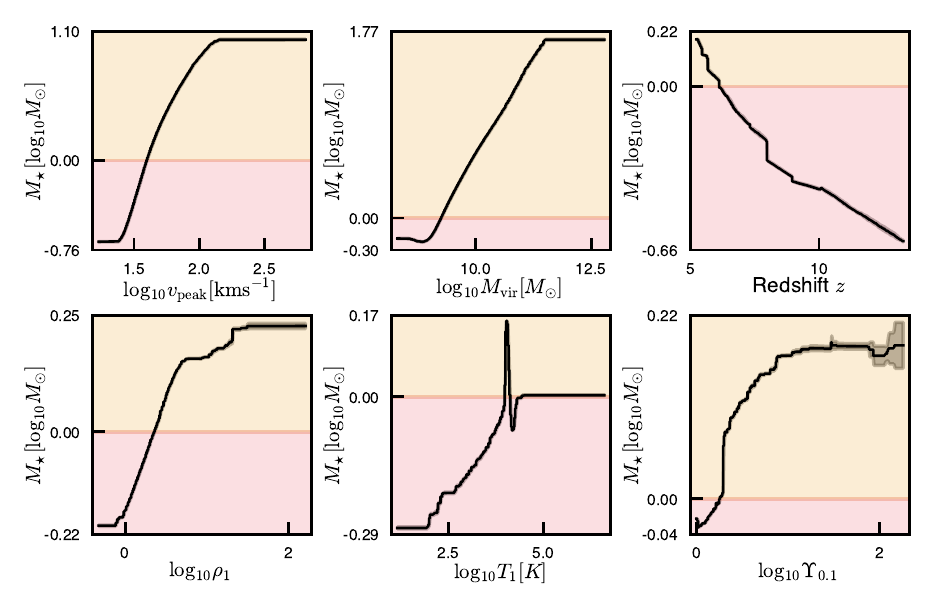}
    \caption{Learned univariate feature functions, $\fyi$ in Equation \ref{eqn:ebm},
             for the EBM $\ebmms$ trained to predict $\mstar$. Areas highlighted in orange
             indicate portions of the function that contribute positively to the predicted
             $\mstar$ and areas in red contribute negatively. Stellar mass increases
             with peak circular velocity and virial mass, increases with decreasing
             redshift, and increases with environmental density. Temperature correlates
             positively with stellar mass, with a strong feature near $\TR\approx10^4$ K
             where hydrogen ionizes. Stellar mass also increases with the mass ratio
             of neighboring halos.}
    \label{fig:mstar_ebm_univariate}
\end{figure*}

The univariate functions determined by the EBM targeting stellar
mass $\mstar$ are shown in Figure
\ref{fig:mstar_ebm_univariate}. Stellar mass increases with
increasing peak circular velocity, virial mass, environmental
density, and neighboring halo mass ratio. Stellar mass increases
with decreasing redshift. As with star formation rate, the
stellar mass increases with increasing environmental temperature
$\TR$, with a sharp enhancement near the temperature where
hydrogen becomes neutral and a sharp deficit near where hydrogen
ionizes.

\subsection{Interaction Functions}

\begin{figure}[h]
    \includegraphics[width=\columnwidth]{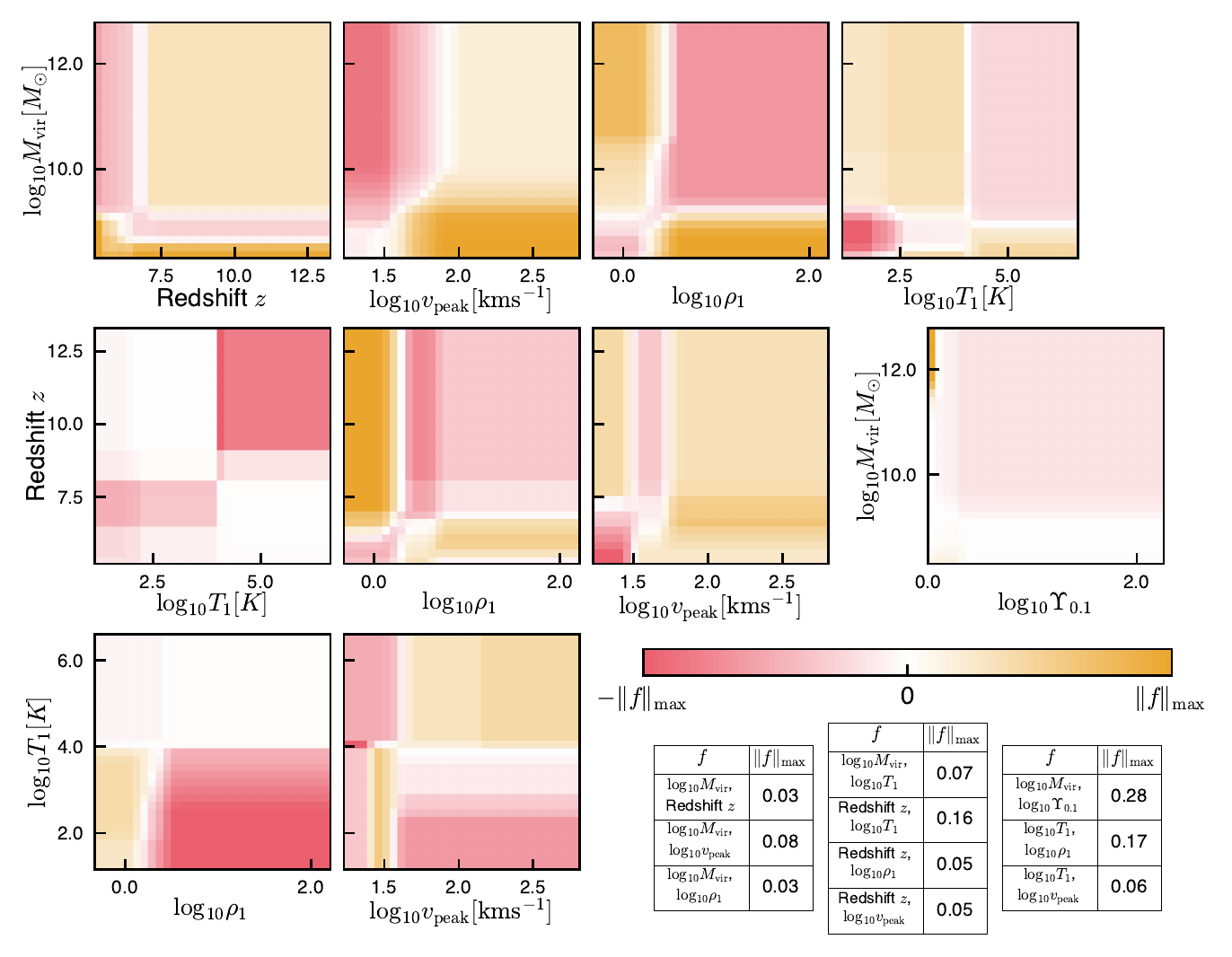}
    \caption{Learned bivariate interaction functions
        $\fyij$ for the EBM $\ebmms$ trained to predict
        $\mstar$. Areas highlighted in orange indicate portions of the
        functions that contribute positively to the predicted $\mstar$ while
        areas in red contribute negatively. Halos with large environmental
        temperatures $\TR$ at high redshift $z$ show enhanced stellar mass.
        The stellar masses of halos with low environmental temperature
        $\TR<10^4$K correlate with environmental density, increasing with
        increasing $\rhoR$. Massive halos with no comparable large neighboring
        halos ($\upR\approx0$) also show enhanced stellar mass.}
    \label{fig:mstar_ebm_interaction}
\end{figure}

The bivariate interaction functions $\fyij$ (see Equation \ref{eqn:ebm})
learned by the EBM when targeting stellar mass $\mstar$
are plotted as heat maps in Figure
\ref{fig:mstar_ebm_interaction}. On average most interaction functions do not
contribute significantly to galaxy stellar mass, but there are
regions of parameter space where the interaction functions are important.
For instance, halos with low environmental temperatures and
high environmental densities have suppressed stellar mass. Large
 virial mass halos with small neighboring halo mass ratios $\log_{10} \upR$,
 indicating
 halos that dominate their local environment, have
  stellar mass enhanced by $\approx 0.3$ dex. This effect
 exceeds the maximum univariate contribution of $\log_{10} \upR$
 alone.
 The deficit of
 stellar mass at environmental temperatures where hydrogen is
 becoming ionized is increased at high redshifts.

\section{Composite EBM}
\label{appendix:cebm}

The composite EBM (CEBM) models we present consist of a \emph{base}
EBM model trained to recover the main trend $\basey$
of the
targeted property $y$ with the input parameters $\vtheta'$,
an \emph{outlier} EBM model that captures the outlying values of $y$
not recovered by $\basey$, and a classification EBM model $\phiy$
that interpolates between them (see \S \ref{sec:cebm}).
Given a CEBM, we
wish to construct analogs of the average contribution, feature
functions, and interaction functions determined for a single
EBM. We define these quantities for the CEBM function in
\S \ref{appendix:cebm_average_contribution} and \ref{appendix:cebm_functions}
below.

\subsection{CEBM Feature and Interaction Functions}

The feature functions of a single EBM are univariate and indicate directly
how the expectation value of the targeted quantity depends on each parameter
$\theta_i\in\vtheta$. With a CEBM comprised of a weighted sum of
two base EBMs, we define the analog of the feature function to be the
weighted sum of the base EBM feature functions. We can write that
\begin{equation}
    \cfeat^i_y = \frac{1}{N}\sum_{j=0}^N \| \phi(\vtheta_j) \odot \bm{f}^i_y(\vtheta_j)\|_{1},
\label{eqn:cebm_feature_function}
\end{equation}
\noindent
where $\odot$ is the
Hadamard or element-wise product operation and the sum is over the number of
samples $N$. The quantity $\bm{f}^i_y$ is the vector of the
individual EBM feature functions $\fyi$.
While the base EBM
feature functions are individually univariate, by weighting the
sum of these feature functions with the classifier EBM the resulting
feature function analog in Equation \ref{eqn:cebm_feature_function} is \emph{not} univariate.

The interaction functions $\tilde{f}^{ij}_y$ are defined
as in Equation \ref{eqn:cebm_feature_function} but with
the vector of the individual EBM interaction functions
$\bm{f}^{ij}_y$ subsituted for $\bm{f}^{i}_y$. While
the interaction functions for a single EBM are
bivariate, the CEBM interaction functions are
\emph{not} bivariate.

\subsection{CEBM Average Contribution}
\label{appendix:cebm_average_contribution}

The average contribution of each feature in a
\cebm{} can be defined
in a manner analogous to the
average contribution computed for a
single EBM (Equation \ref{eqn:average_contribution}).
The CEBM average contribution can be written as
\begin{equation}
    \bar{f}_y^i = \frac{\sum_{j=0}^{n_b-1} \tilde{f}(\theta_{i,j}) N_j}{\sum_{j=0}^{n_b-1}N_j}
\label{eqn:cebm_average_contribution}
\end{equation}
\noindent
where $\cfeat$ is either the CEBM feature function $\cfeat^i_y$
or the CEBM interaction function $\cfeat^{ij}_y$.
Equation \ref{eqn:cebm_average_contribution} characterizes
how important the parameter $\theta_i$ is for modeling
the target quantity $y$.

\subsection{Visualizing CEBM Feature and Interaction Functions}
\label{appendix:cebm_functions}

The feature and interaction functions $\cfeat^i_y$ and $\cfeat^{ij}_y$ are not
univariate or bivariate by design, which allows them to model the
outlier distribution about the base EBM model $\basey$.
To visualize the feature and interaction functions for CEBM models
in a manner similar to the univariate feature and bivariate interaction
functions for single EBM, we can average the values of
$\bm{f}^i_y$ and $\bm{f}^{ij}_y$.
For the feature function averaged over $N$ samples, consider
$n_b$ bins along the $\theta_i$ direction, with central
values $\theta_{i,b}$ and bin widths
$\Delta \theta_{i,b}$.
The bin-averaged CEBM feature and interaction functions are then
\begin{equation}
    f^{i,b}_y = \frac{1}{N} \sum_{j=0}^{N-1} \alpha(\theta_{i,b},  \Delta  \theta_{i,b},\theta_{j,i})\phi(\vtheta_j) \odot \bm{f}(\vtheta_j)
\label{eqn:cebm_features_visualization}
\end{equation}
\noindent
where $\theta_{j,i}$ is the $i$th parameter of the $j$th sample $\vtheta_j$ and
the function $\alpha(\theta_{i,b},  \Delta  \theta_{i,b},\theta_{j,i})=1$ if $\theta_{i,b}-\Delta \theta_{i,b}/2 \leq \theta_{j,i} \leq \theta_{i,b}+\Delta \theta_{i,b}/2$ and $\alpha=0$ otherwise.
The quantity $\bm{f}$ is either the vector of EBM feature functions $\bm{f}^i_y$
or the EBM interaction functions $\bm{f}^{ij}_y$.
Equation
\ref{eqn:cebm_features_visualization} calculates the mean of the $\bm{f}$ values in each
of the $n_b$ bins, and can be modified to calculate its standard deviation.

\section{Composite EBM Model for Star Formation Rate}
\label{appendix:cebm_sfr}

The CEBM model $\cebmsfr$ for the
star formation rate consists of a base EBM $\basesfr$,
a residual EBM $\outsfr$ that attempts to capture the outlying values of $\sfr$
not recovered by $\basesfr$, and the classifier EBM
$\phisfr$. For each of these individual EBMs that form the CEBM model,
we plot the average contribution, feature functions, and interaction functions.

\begin{figure}
    \centering
    \includegraphics[width=\textwidth]{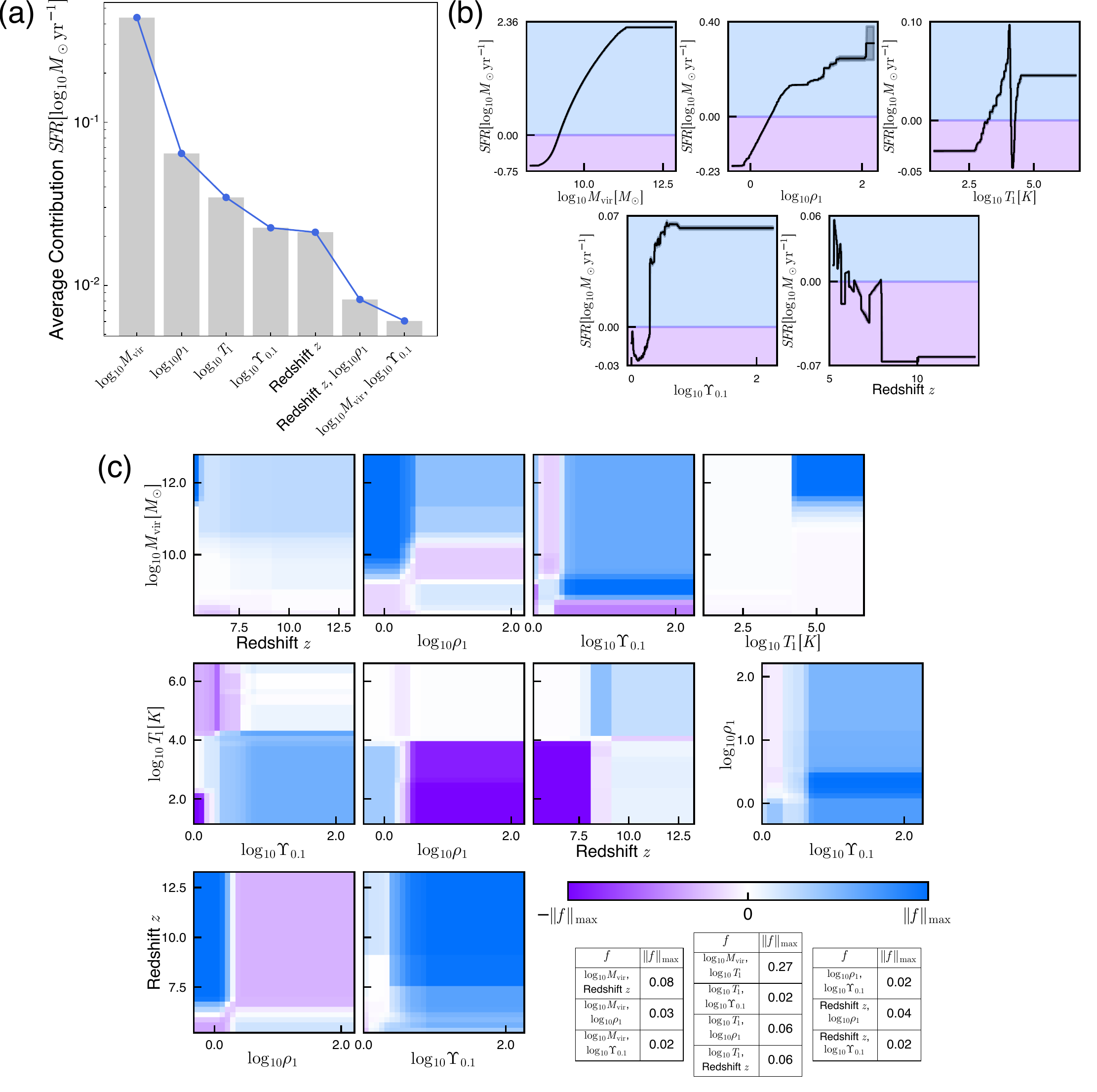}
    \caption{Details for the base EBM model $\basesfr$ component
    of the CEBM $\cebmsfr$ trained to predict $\sfr$.
    Panel a) displays the
    average contribution of features.
    The dominant feature is virial mass $\mvir$, with an
    average contribution to $\log_{10} \sfr$ roughly $8-10\times$ larger than
    environmental density $\rhoR$ and temperature $\TR$. Compared with the
    average contributions to the EBM $\ebmsfr$ (see Figure \ref{fig:sfr_ebm_avg_cont}),
    $\mvir$ subsumes the contribution provided by the missing $\vpeak$ parameter.
    Panel b) shows the
    feature functions contributing to the base EBM model. The SFR increases
    with $\mvir$, which provides the dominant contribution. A secondary
    contribution comes from environmental density $\rhoR$. Environmental
    temperature $\TR$ has a minor contribution, but shows the same
    feature at $\TR\approx10^{4}$K where hydrogen ionizes. The mass
    ratio of nearby halos $\upR$ and redshift $z$ provide minor contributions.
    Panel c) presents the
    interaction functions for the base EBM $\basesfr$. Each panel
    shows the contribution of the bivariate interaction terms, normalized
    such that the color map ranges between plus or minus the maximum of
    the norm of each function $||f||_\mathrm{max}$. Purple indicates
    negative contributions and blue indicates positive
    contributions.
    The table lists
    $||f||_\mathrm{max}$ for the interaction functions, each with
    units $\log_{10}\Msun~\yr^{-1}$. In absolute terms, the largest
    interaction occurs for large virial mass $\mvir$ and environmental
    temperature $\TR$. SFR is partially reduced for low environmental
    temperature $\TR$ and either low environmental density $\rhoR$ or
    redshift $z$.
    }
    \label{fig:sfr_base_ebm}
\end{figure}

Figure \ref{fig:sfr_base_ebm} shows the average contribution, feature
functions, and interaction functions for the EBM model
$\basesfr${}
that forms the base of the CEBM model. The differences between
$\ebmsfr$ and $\basesfr$ reflect the additional information
provided by the maximum peak circular velocity $\vpeak$. Without
access to $\vpeak$, the base EBM $\basesfr$ upweights $\bar{f}(\mvir)$
such that its importance roughly equals the combined importance of $\mvir$
and $\vpeak$ in determining $\ebmsfr$. The average contribution
of $\rhoR$, $\TR$, $z$, $\upR$, and $(\mvir,\upR)$ are similar between the models.
The additional interaction term in the top seven average contributions
is $(z,\rhoR)$, with a percent-level contribution to $\sfr$ relative to $\mvir$.
The feature functions for $\basesfr$ have shapes similar to the
feature functions for $\ebmsfr$, but their minimum and maximum contributions
to $\sfr$ are adjusted to account for the missing $\vpeak$ contribution. The
feature function $\bar{f}(z)$ is noisier overall. For the interaction functions,
the largest contributors now involve $\mvir$ rather than the missing parameter $\vpeak$,
and the set of available functions is substantially different than with $\ebmsfr$.

\begin{figure}
    \centering
    \includegraphics[width=\textwidth]{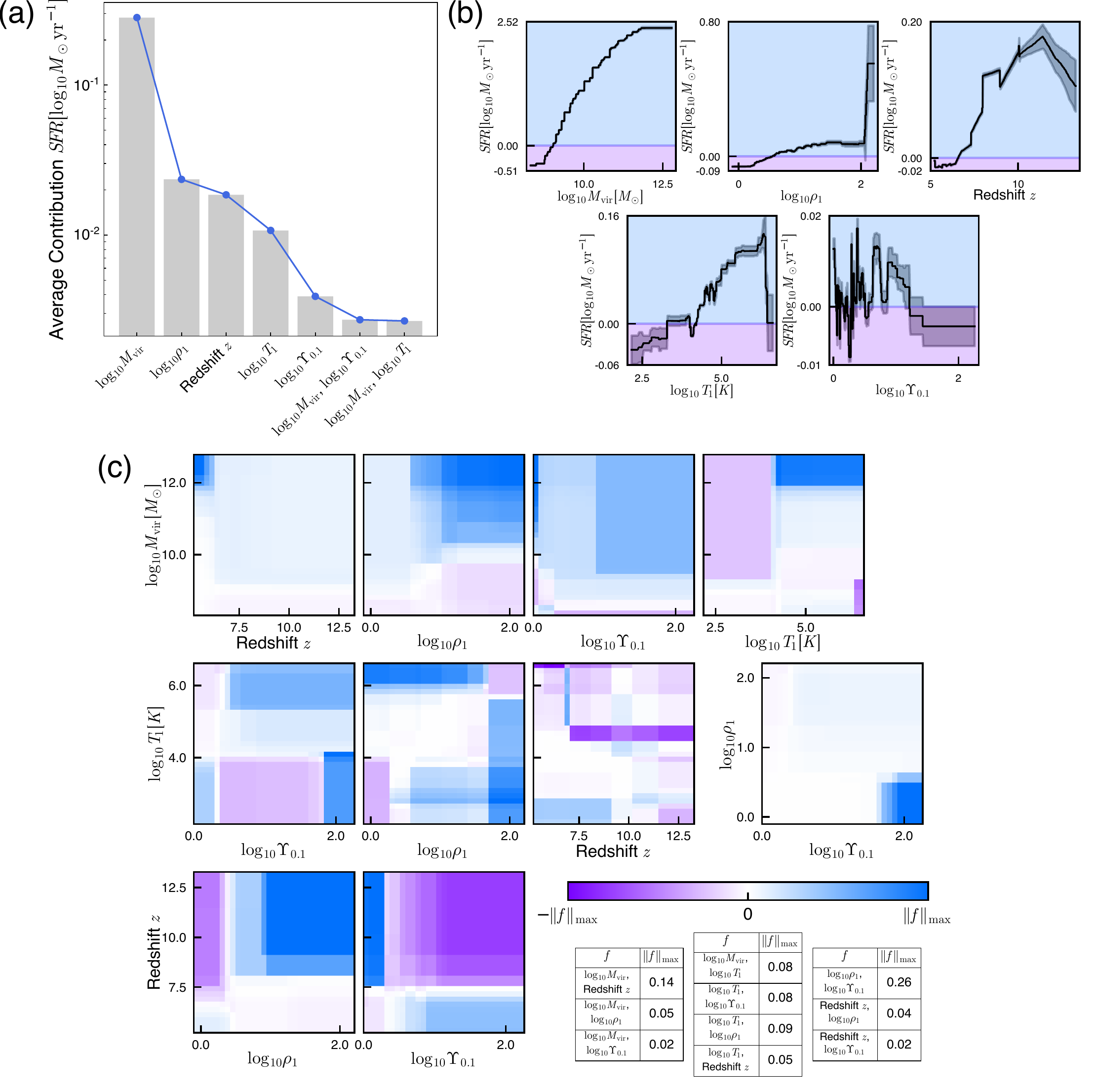}
    \caption{Details for the outlier EBM model $\outsfr$ component
    of the CEBM $\cebmsfr$ trained to predict $\sfr$.
    Panel a) displays the
    average contribution of features. As with the base EBM $\basesfr$, the
    feature with the largest average contribution
    is virial mass $\mvir$, with roughly $\gtrsim10\times$ larger
    contribution to $\log_{10}\sfr$ than
    environmental density $\rhoR$, redshift $z$, or temperature $\TR$. The average
    contributions of $\upR$ or interactions are small.
    Panel b) shows the
    feature functions for the outlier EBM $\outsfr$. The feature
    function for virial mass $\mvir$ has the largest contribution to
    $\outsfr$, similar to the virial mass dependence of the base EBM $\basesfr$
    (see Panel b) of Figure \ref{fig:sfr_base_ebm}). The SFR of outliers increases
    with increasing environmental density $\rhoR$, with a large enhancement
    at very large $\rhoR$. Unlike the base EBM $\basesfr$, SFR for the outliers
    increases with increasing redshift. The feature function for the nearby
    halo mass ratio $\upR$ is weak and noisy.
    Panel c) presents the
    interaction functions for the outlier EBM $\outsfr$. Each panel
    shows the contribution of the bivariate interaction terms, normalized
    such that the color map ranges between plus or minus the maximum of
    the norm of each function $||f||_\mathrm{max}$. Purple indicates
    negative contributions and blue indicates positive
    contributions.
    The table lists
    $||f||_\mathrm{max}$ for the interaction functions, each with
    units $\log_{10}\Msun~\yr^{-1}$. For outliers, the SFR increases
    at low environmental density $\rhoR$ and large neighboring halo
    mass ratios $\upR$, suggesting dynamical interactions increase
    star formation rate in low density environments. The other interaction
    functions are relatively weak.
    }
    \label{fig:sfr_outlier_ebm}
\end{figure}

Figure \ref{fig:sfr_outlier_ebm}
shows the
average contribution, feature functions, and
interaction functions for the outlier EBM $\outsfr$ fit to
the deviant samples not captured by the base EBM $\basesfr$.
The outlier EBM receives the highest contribution from virial
mass, with an average contribution more than an order of
magnitude larger than the next most important feature $\rhoR$.
The redshift $z$ and environmental temperature $\TR$ have
comparable importance to $\rhoR$. The remaining features
provide only percent-level contributions relative to $\mvir$.

\begin{figure}
    \centering
    \includegraphics[width=\textwidth]{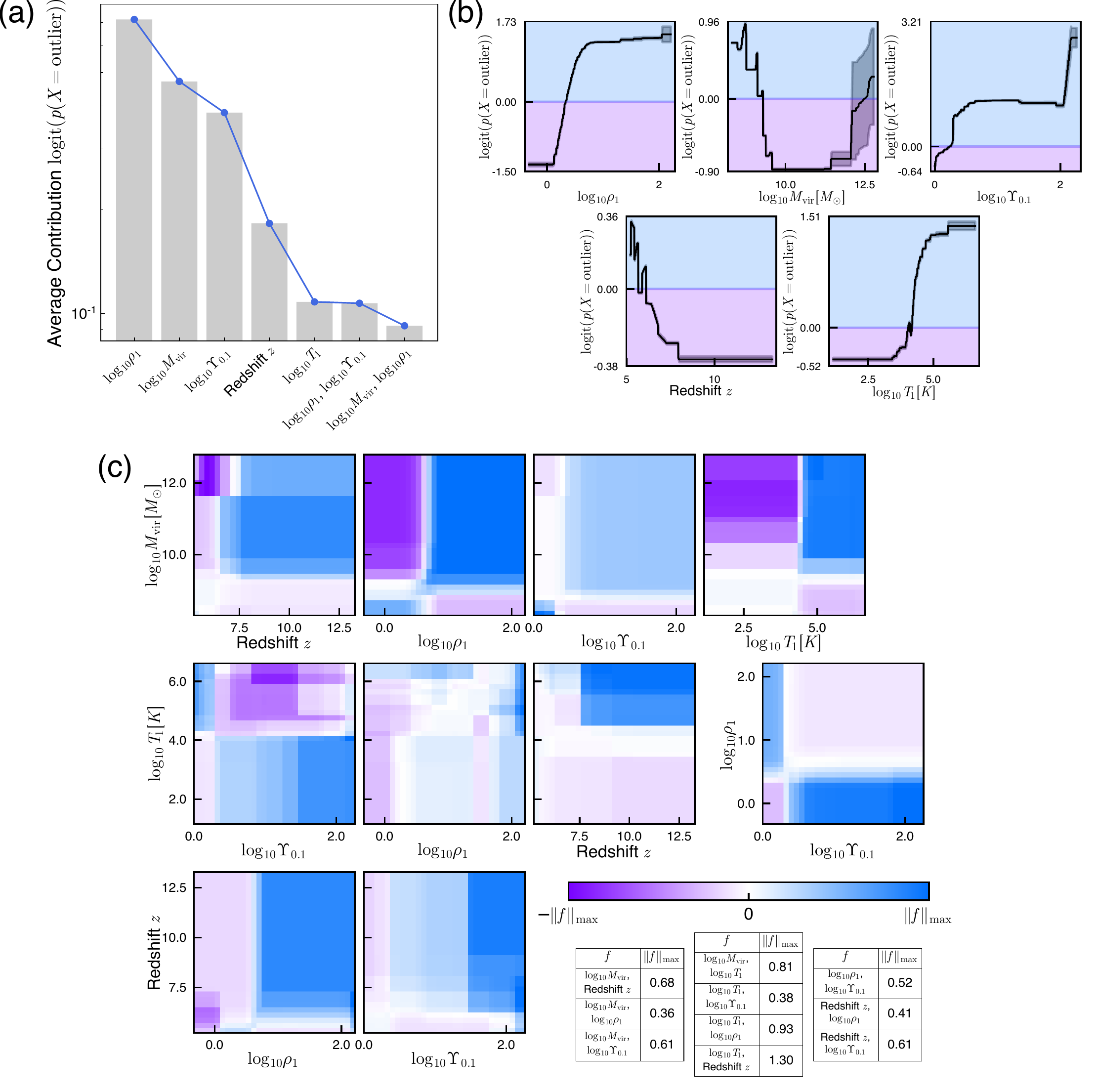}
    \caption{Details for the classification EBM model $\classsfr$
    that interpolates between
    the base EBM $\basesfr$ and the outlier EBM $\outsfr$ for creating
    the CEBM $\cebmsfr$.
    Panel a) displays the
    average contribution of features to the
    classification EBM model $\classsfr$.
    The most important features for determining
    whether a galaxy is an outlier in the SFR distribution are
    environmental density $\rhoR$, virial mass $\mvir$, and nearby
    halo mass ratio $\upR$. The average contributions are unit free,
    and represent changes to the log odds of a galaxy being an outlier
    in the SFR distribution.
    Panel b) shows the
    feature functions contributing to the classifier EBM $\classsfr$. These feature functions represent the
    change in log odds that a given galaxy will be an outlier in SFR.
    Outliers
    tend to occur at high environmental density $\rhoR$ or
    very low or high virial masses $\mvir$. Galaxies with massive neighbors,
    reflected by $\upR$,
    or high environmental temperature $\TR$ are also more likely to be outliers.
    The lowest redshift galaxies in the dataset are additionally likely be outliers
    in SFR.
    Panel c) presents the
    interaction functions for the classifier EBM $\classsfr$. Each panel
    shows the contributions of the interaction terms, normalized
    such that the color map ranges between plus or minus the maximum of
    the norm of each function $||f||_\mathrm{max}$. Purple indicates
    negative log odds and blue indicates positive
    log odds that a given galaxy is an outlier in SFR.
    The table lists
    $||f||_\mathrm{max}$ for the interaction functions, listed as
    the corresponding change in log odds. Galaxies with large
    environmental temperature $\TR$ and with high environmental
    density $\rhoR$, at high redshift $z$, or with large virial
    mass $\mvir$ are more likely to be outliers. Massive galaxies
    in high environmental densities or at high redshift also
    are more likely outliers in SFR.
    }
    \label{fig:sfr_classifier_ebm}
\end{figure}

Figure \ref{fig:sfr_classifier_ebm}
shows the
average contribution, feature functions, and
interaction functions for the classifier EBM $\outsfr$
that interpolates between the base and outlier
EBMs when calculating the CEBM model.
For the classifier EBM, the most important
features are $\rhoR$, $\mvir$, and $\upR$.
Redshift $z$ has middling importance, following
by $\TR$, $[\rhoR, \upR]$, and $[\mvir, \rhoR]$.
The feature functions show strong dependencies on
$\rhoR$, $\mvir$, $\upR$, $z$, and $\TR$.
The largest interaction functions involve
the environmental temperature $\TR$, redshift
$z$, and virial mass $\mvir$.

\begin{figure}[h]
    \centering
    \includegraphics[width=\textwidth]{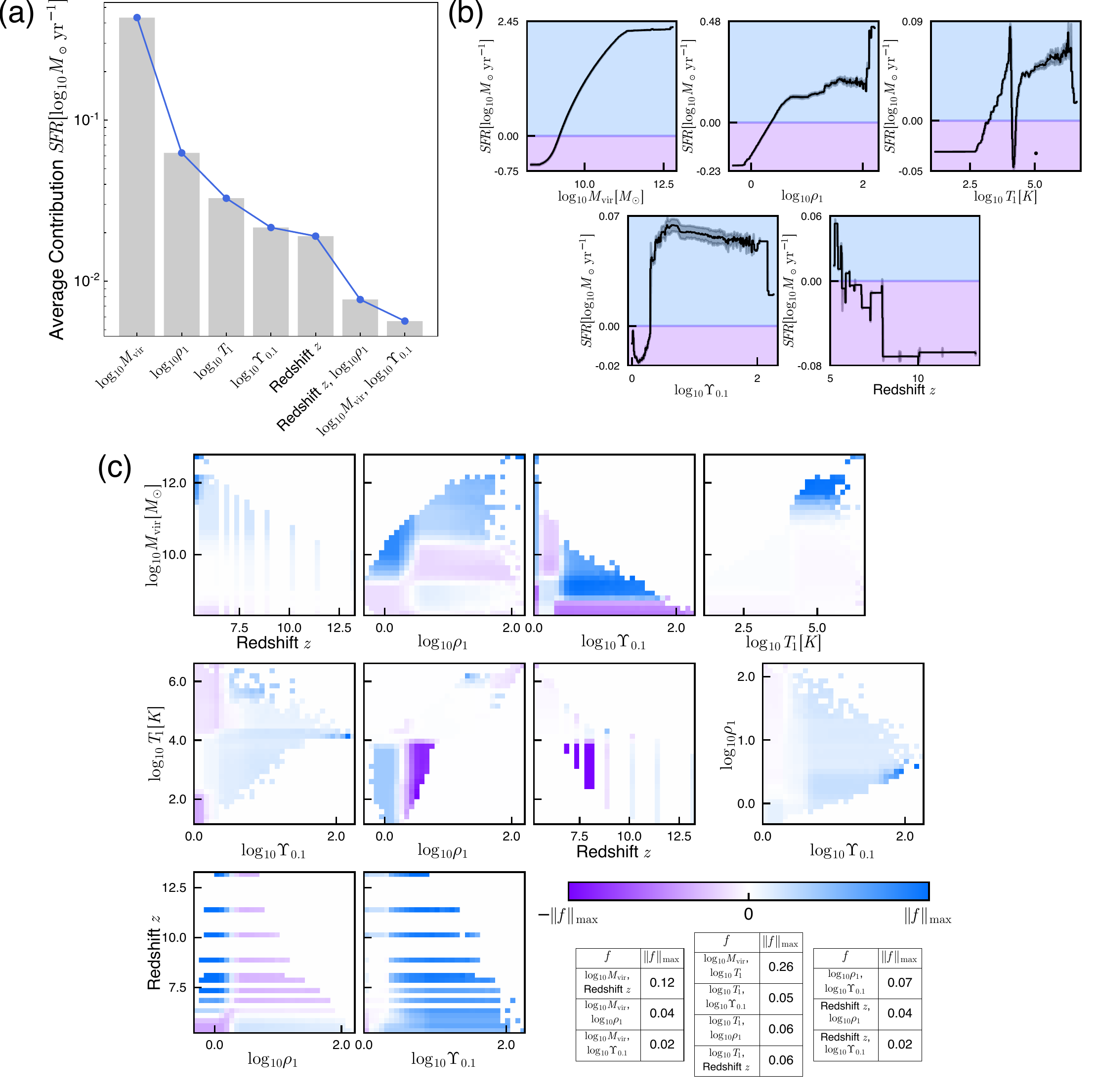}
    \caption{Details for the CEBM model $\cebmsfr$
    trained to predict $\sfr$.
    Panel a) displays the
    average contribution of features to the CEBM.
    Virial mass $\mvir$ provides the largest
    average contribution to the star formation rate. The environmental
    density $\rhoR$ provides a $\sim6\times$ smaller average contribution. The
    environmental temperature $\TR$, nearby galaxy mass ratio $\upR$,
    and redshift $z$ provide a relative contribution roughly $10\times$
    smaller than $\mvir$.
    Panel b) shows the
    feature functions contributing to the
    CEBM $\cebmsfr$. The SFR increases
    with $\mvir$, which provides the largest contribution. A secondary
    contribution comes from environmental density $\rhoR$. Environmental
    temperature $\TR$ has a minor contribution, but shows the familiar
    feature at $\TR\approx10^{4}$K where hydrogen ionizes. The mass
    ratio of nearby halos $\upR$ and redshift $z$ provide minor contributions.
    As expected, the CEBM feature functions are similar to the base EBM
    feature functions that represent the parameter dependence of star formation
    rate for most galaxies in the dataset (see Panel b) of Figure \ref{fig:sfr_base_ebm}).
    Panel c) presents the
    interaction functions for the CEBM $\cebmsfr$. Each panel
    shows the contribution of the interaction terms, normalized
    such that the color map ranges between plus or minus the maximum of
    the norm of each function $||f||_\mathrm{max}$. Purple indicates
    negative contributions and blue indicates positive
    contributions.
    The table lists
    $||f||_\mathrm{max}$ for the interaction functions, each with
    units $\log_{10}\Msun~\yr^{-1}$. As for the interaction functions
    for the base EBM $\basesfr$, the largest
    interaction occurs for large virial mass $\mvir$ and large environmental
    temperature $\TR$ or low redshift $z$.
    }
    \label{fig:sfr_composite_ebm}
\end{figure}

By weighting the base and outlier EBM models
with the classifier EBM, we construct the
CEBM for star formation rate as
$\cebmsfr \equiv [1-\classsfr]\basesfr + \classsfr\outsfr$.
Figures \ref{fig:sfr_composite_ebm}
show the average contribution, feature functions,
and interaction functions for the $\sfr$ CEBM.
The most important feature is $\mvir$, which
dominates by a factor of $\sim4-10$ over
environmental density $\rhoR$, environmental
temperature $\TR$, $\upR$, and redshift $z$.
The interaction terms are roughly percent-level
effects relative to $\mvir$. The feature
functions show a strongly increasing $\sfr$
with $\mvir$, and enhanced $\sfr$ with
environmental density $\rhoR$.
The temperature dependence shows the feature
at $\log_{10}\TR \approx 4$ seen with the
EBM model $\ebmsfr$.
The interaction functions provide only important
contributions over very limited areas of parameter space,
with the most important adjustments occuring at low
redshift and large virial mass, or for large temperatures
and virial masses.
For reference, the model
summary Figure \ref{fig:sfr_cebm_model_summary}
illustrates the overall performance of the model.

\section{Composite EBM Model for Stellar Mass}
\label{appendix:cebm_mstar}

The CEBM model $\cebmms$ for stellar mass
is comprised of a base EBM $\basems$,
an outlier EBM that attempts to model the $\mstar$ of samples
not recovered by $\basems$, and the classifier EBM function
$\phims$ that interpolates between them. The average contribution, feature functions, and
interaction functions from these component EBM models are presented below.

Figures \ref{fig:mstar_base_ebm} shows the average contribution, feature
functions, and interaction functions for the base EBM model $\basems$.
By removing $\vpeak$ from the dataset used to train the EBM, the
base EBM model for the $\mstar$ CEBM replaces the dependence on
$\vpeak$ with an additional dependence on $\mvir$. The relative ordering
and importance of redshift $z$, environmental density $\rhoR$,
environmental temperature $\TR$, and $\upR$ are approximately maintained.
For the feature functions, the results shown for $\basems$ in Panel b) of
Figure
\ref{fig:mstar_base_ebm} can be compared with
the results for $\ebmms$ shown in Figure \ref{fig:mstar_ebm_avg_cont}.
As reflected by average contributions, the amplitude of
the feature function $\bar{f}(\mvir)$ increases to account for the
removal of $\vpeak$. The feature functions for $z$, $\rhoR$, $\TR$,
and $\upR$ are modified and remain similar in shape to those
computed for the EBM $\ebmms$.
The interaction functions shared between $\basems$ and $\ebmms$ are
similar. There is an increase in $\mstar$ contribution for large
$[\mvir,\TR]$ and a decrease in the amplitude of $[\mvir,\upR]$.

Figure \ref{fig:mstar_outlier_ebm}
shows the
average contribution, feature functions, and
interaction functions for the outlier EBM model $\outms$.
The average contribution is dominated by $\mvir$, with
the contributions from all other single parameters lower by
a factor of $\approx10$ with the order of importance
maintained relative to $\basems$.
For the feature functions, the redshift dependence
changes dramatically and now increases with increasing
redshift. The feature function for environmental
density $\cfeat(\rhoR)$ becomes much weaker over
a wide range of $\rhoR$, but increases dramatically
at high $\rhoR$. Relative to the $\basems$ feature functions,
the feature function $\cfeat(\upR)$ for $\outms$ is weak
and noisy.
The interaction functions show increased
contributions at large $[z,\rhoR]$, and for
low $\TR$ and large $\rhoR$.

\begin{figure}
    \centering
    \includegraphics[width=\textwidth]{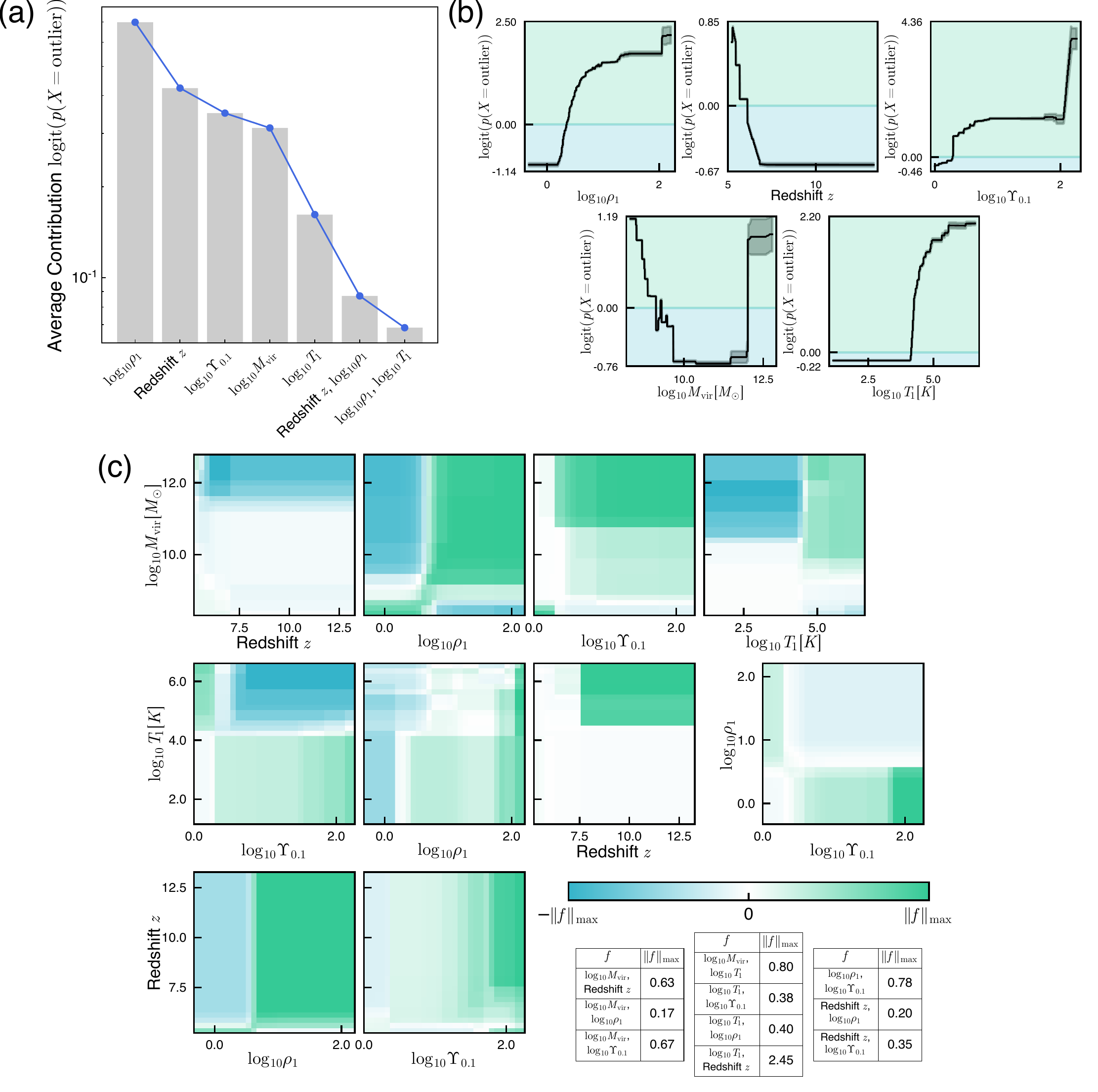}
    \caption{Details for the classification EBM model $\classms$
    that interpolates between
    the base EBM $\basems$ and the outlier EBM $\outms$ for creating
    the CEBM $\cebmms$.
    Panel a) displays the
    average contribution of features to the
    classification EBM model $\classms$. The most important features for determining
    whether a galaxy is an outlier in the stellar mass distribution are
    environmental density $\rhoR$, redshift $z$, nearby
    halo mass ratio $\upR$, and
    virial mass $\mvir$. The average contributions are unit free,
    and represent changes to the log odds of a galaxy being an outlier
    in the stellar mass distribution.
    Panel b) shows the
    feature functions contributing to the classifier EBM $\classms$.
    These feature functions represent the
    change in log odds that a given galaxy will be an outlier in $\mstar$. Outliers
    tend to occur at high environmental density $\rhoR$ or
    very low or high virial masses $\mvir$. Galaxies with massive neighbors,
    reflected by $\upR$,
    or high environmental temperature $\TR$ are also more likely to be outliers.
    The lowest redshift galaxies in the dataset are additionally likely be outliers
    in stellar mass. These trends are similar to the feature functions for the
    classifier EBM $\classsfr$ (see Panel b) of Figure \ref{fig:sfr_classifier_ebm}).
    Panel c) presents the
    interaction functions for the classifier EBM $\classms$. Each panel
    shows the contributions of the interaction terms, normalized
    such that the color map ranges between plus or minus the maximum of
    the norm of each function $||f||_\mathrm{max}$. Teal indicates
    negative log odds and green indicates positive
    log odds that a given galaxy is an outlier in stellar mass.
    The table lists
    $||f||_\mathrm{max}$ for the interaction functions, listed as
    the corresponding change in log odds. Galaxies with large
    environmental temperature $\TR$ and at high redshift $z$ are
    more likely to be outliers. Massive galaxies at high environmental
    density $\rhoR$ or with large nearby halos (large $\upR$) also
    tend to be outliers. Galaxies at low environmental density but
    with large nearby halos are also have an increased likelihood
    of being outliers in stellar mass.
    }
    \label{fig:mstar_classifier_ebm}
\end{figure}

Figure \ref{fig:mstar_classifier_ebm}
shows the
average contribution, feature functions, and
interaction functions for the classifier EBM
$\classms$. For each of these properties, we
note that in determining $\classms$ a
sigmoid function $\sigma$ is applied to the
sum of $\beta$, $\fyi$ and $\fyij$ that
model the log odds that a galaxy is an outlier
in stellar mass.
In determining $\mstar$, the features with
the largest average contribution
are environmental density $\rhoR$,
redshift $z$, $\upR$, and virial mass
$\mvir$. The interaction terms
with the largest contribution are
$(z,\rhoR)$ and $(\rhoR,\TR)$. Clearly,
environmental density plays an important
role in determining whether a given
simulated galaxy is an outlier relative
to the base EBM $\basems$.
The feature functions show that galaxies
with large environmental densities $\rhoR$,
at low redshift $z$, or with a large neighboring
galaxy (expressed by $\upR$) have an enhanced
probability of being
outliers relative to $\basems$.
Galaxies at both high and low $\mvir$ or large
environmental temperature $\TR$ are also more
likely to be outliers.

We construct the stellar mass CEBM with the
sum $\cebmms \equiv [1-\classms]\basems + \classms\outms$.
Figure \ref{fig:mstar_composite_ebm}
shows the average contribution, feature functions,
and interaction functions for $\cebmms$.
The feature with largest average
contribution is $\mvir$, with redshift $z$,
environmental density $\rhoR$, environmental
temperature $\TR$, and the mass ratio of
nearby galaxies $\upR$ having an
lower average contribution by a factor of $\sim5-10$.
Relative to $\mvir$, the interactions
$[z,\rhoR]$ and $[\mvir,\rhoR]$ contribute at
level of a few percent.
The $\mstar$ CEBM feature function $\cfeat(\mvir)$
has increased in amplitude relative to the $\mstar$ EBM
feature function
$f(\mvir)$, subsuming some of the dependence on the
missing $\vpeak$ feature.
The remaining feature functions for $\cebmms$ are similar
in shape and amplitude to those for $\ebmms$, although
the contribution at large $\TR$ and $\rhoR$ are increased
and the dependence on redshift $z$ is decreased.
The interaction functions are similar between $\cebmms$
and $\basems$, although there is a larger enhancement of $\mstar$
for large $[\mvir,\TR]$ and a smaller enhancement for
large $\mvir$ and small $\upR$ for the CEBM $\cebmms$.
For reference, the model
summary Figure \ref{fig:mstar_cebm_model_summary}
illustrates the overall performance of the model.

\begin{figure}
    \centering
    \includegraphics[width=\textwidth]{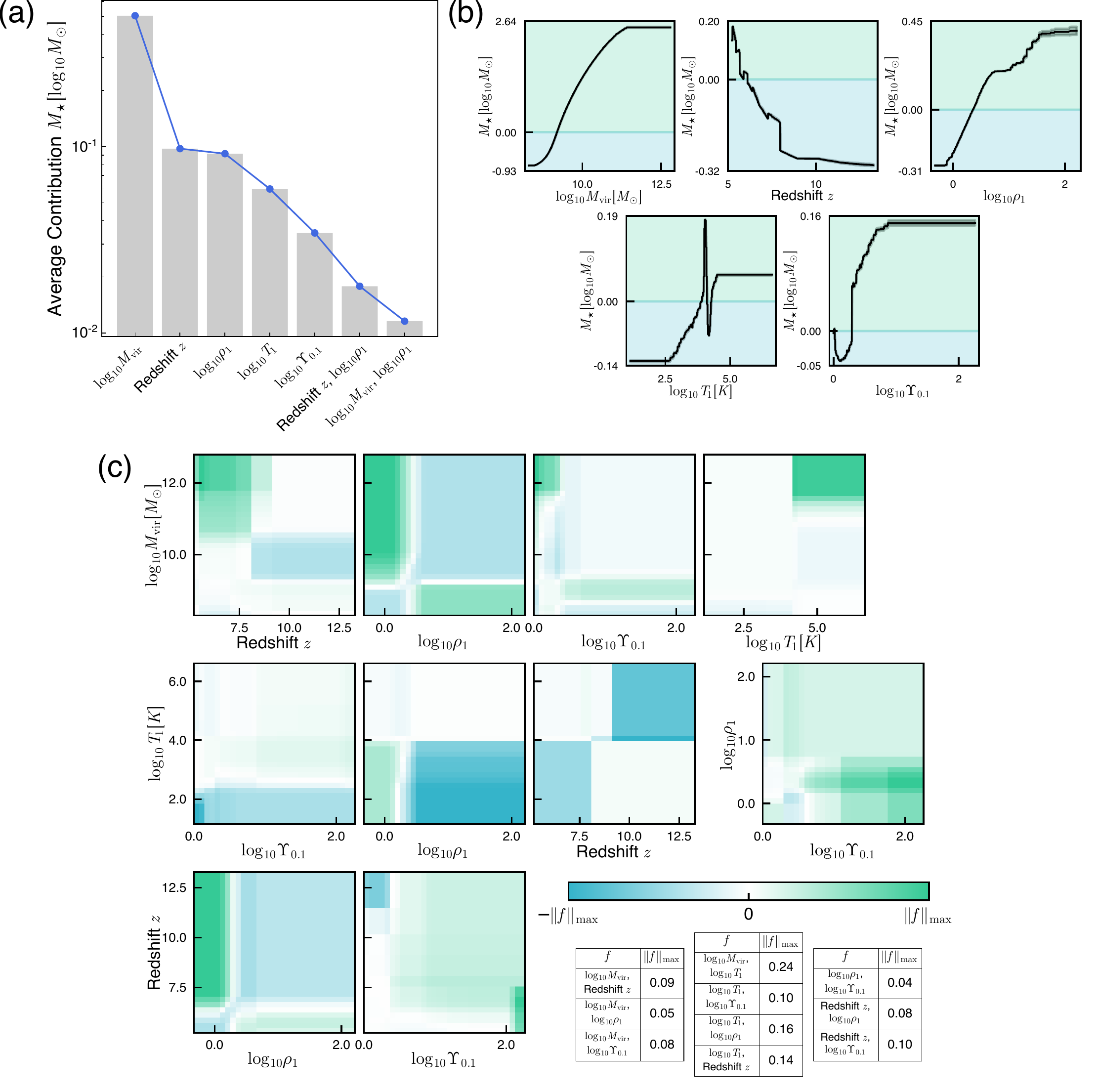}
    \caption{Details for the base EBM model $\basems$ component
    of the CEBM $\cebmms$ trained to predict stellar mass $\mstar$.
    Panel a) displays the
    average contribution of features to the base EBM model $\basems$.
    The feature with the highest average contribution is
    virial mass $\mvir$, with an
    average contribution to $\log_{10} \mstar$ roughly $5\times$ larger than
    that from redshift $z$ or
    environmental density $\rhoR$. The environmental temperature $\TR$
    and nearby halo mass ratio $\upR$ provide smaller average contributions,
    and interactions between features are yet smaller.
    Panel b) shows the
    feature functions contributing to the base EBM model $\basems$. The
    stellar mass increases
    with $\mvir$ that provides the largest contribution. Secondary
    contributions come from redshift $z$, which increases $\mstar$ at
    later times, and the positive correlate environmental density $\rhoR$.
     comes from environmental density $\rhoR$. Environmental
    temperature $\TR$ has a small contribution, and shows the familiar
    feature at $\TR\approx10^{4}$K where hydrogen ionizes. The mass
    ratio of nearby halos $\upR$ provides a minor contribution.
    Panel c) presents the
    interaction functions for the base EBM $\basems$. Each panel
    shows the contribution of the bivariate interaction terms, normalized
    such that the color map ranges between plus or minus the maximum of
    the norm of each function $||f||_\mathrm{max}$. Teal indicates
    negative contributions and green indicates positive
    contributions.
    The table lists
    $||f||_\mathrm{max}$ for the interaction functions, each with
    units $\log_{10}\Msun$. In absolute terms, the largest
    interaction occurs for large virial mass $\mvir$ and environmental
    temperature $\TR$ (same as for the base EBM $\basesfr$ modeling
    star formation rate, see Panel c) of Figure \ref{fig:sfr_base_ebm}).
    Stellar mass is partially reduced for low environmental
    temperature $\TR$ and either high environmental density $\rhoR$ or
    high redshift $z$.
    }
    \label{fig:mstar_base_ebm}
\end{figure}

\begin{figure}
    \centering
    \includegraphics[width=\textwidth]{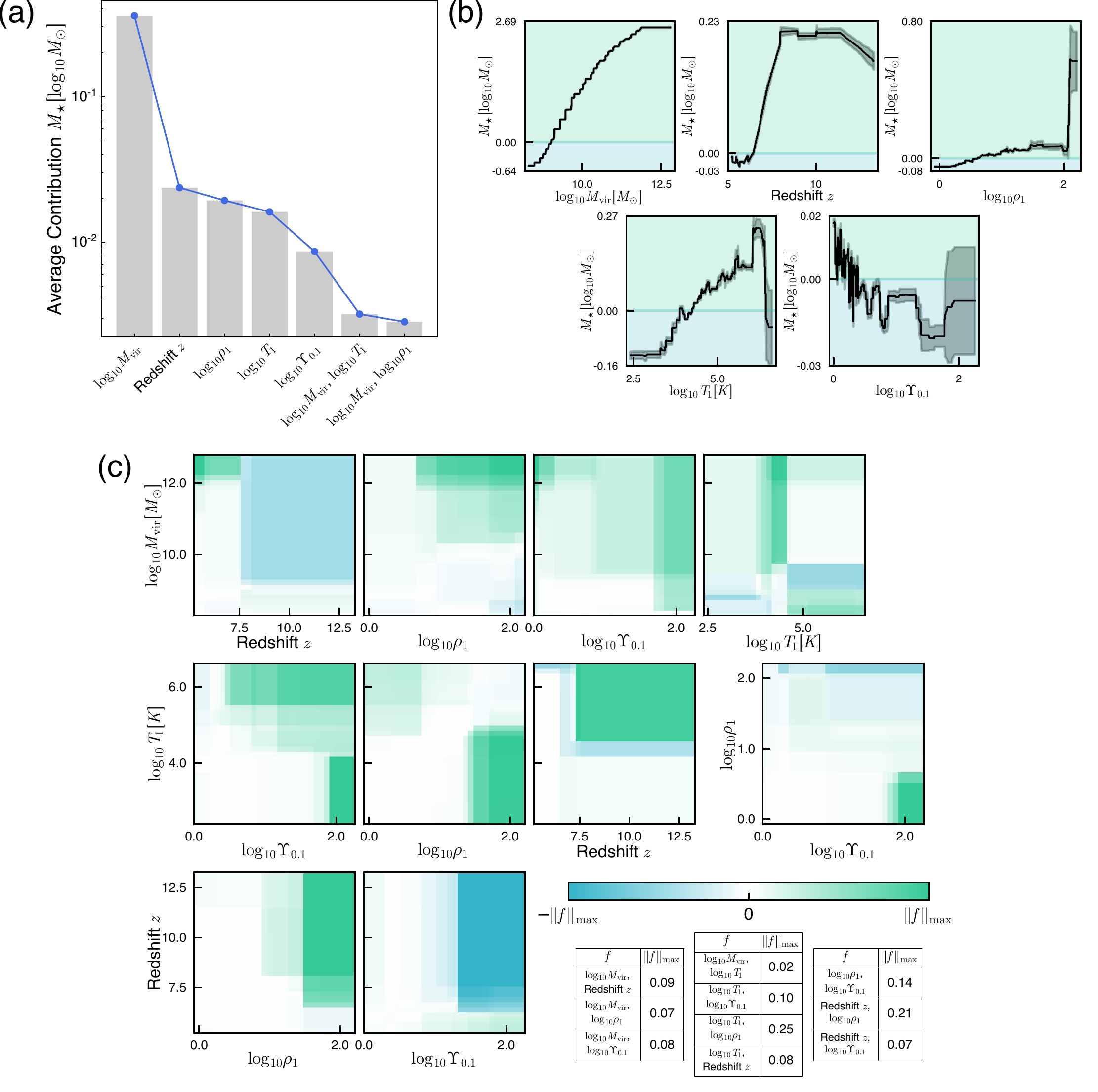}
    \caption{Details for the outlier EBM model $\outms$ component
    of the CEBM $\cebmms$ trained to predict $\mstar$.
    Panel a) displays the
    average contribution of features to the outlier EBM model $\outms$.
    As with the base EBM $\basems$, the
    feature with the largest average contribution
    is virial mass $\mvir$, with roughly $\gtrsim10\times$ larger
    contribution to $\log_{10}\mstar$ than redshift $z$,
    environmental density $\rhoR$, or temperature $\TR$. The average
    contributions of $\upR$ or interactions are small.
    Panel b) shows the
    feature functions for the outlier EBM $\outms$. The feature
    function for virial mass $\mvir$ has the largest contribution to
    $\outsfr$, similar to the virial mass dependence of the base EBM $\basems$
    (see Panel b) of Figure \ref{fig:mstar_base_ebm}).
    The stellar mass of outliers increases
    with increasing environmental density $\rhoR$, with a large enhancement
    at very large $\rhoR$. Unlike the base EBM $\basems$, the stellar mass of
    the outliers
    increases with increasing redshift. The feature function for the nearby
    halo mass ratio $\upR$ is weak and noisy.
    Panel c) presents the
    interaction functions for the outlier EBM $\outms$. Each panel
    shows the contribution of the interaction terms, normalized
    such that the color map ranges between plus or minus the maximum of
    the norm of each function $||f||_\mathrm{max}$. Teal indicates
    negative contributions and green indicates positive
    contributions.
    The table lists
    $||f||_\mathrm{max}$ for the interaction functions, each with
    units $\log_{10}\Msun$. For outliers, stellar mass increases
    at high environmental density $\rhoR$
    with low environmental temperature $\TR$ or high redshift $z$.
    }
    \label{fig:mstar_outlier_ebm}
\end{figure}

\begin{figure}
    \centering
    \includegraphics[width=\textwidth]{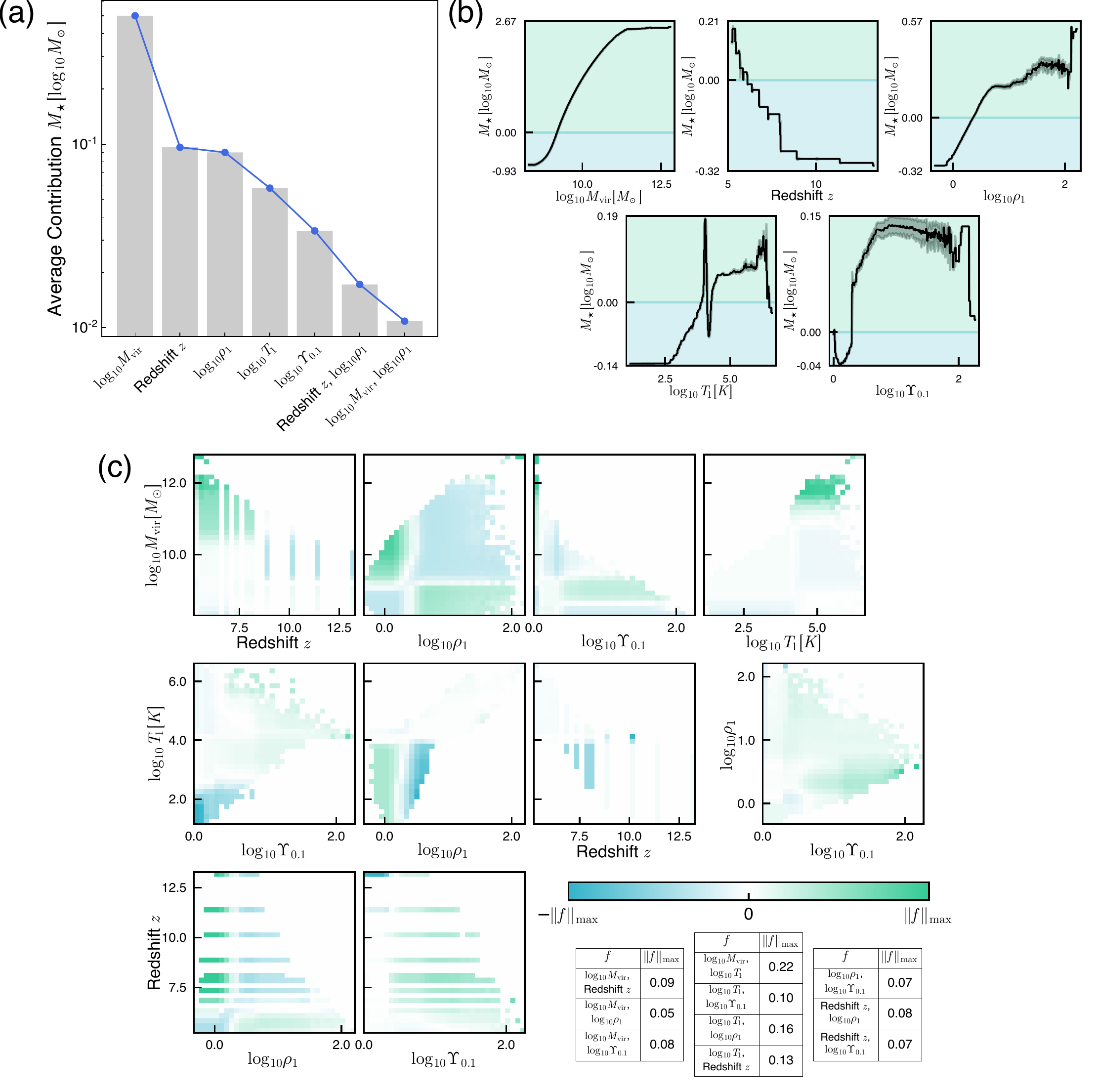}
    \caption{Details for the CEBM model $\cebmms$
    trained to predict stellar mass $\mstar$.
    Panel a) displays the
    average contribution of features to the CEBM model $\cebmms$.
    Virial mass $\mvir$ provides the largest
    average contribution to the stellar mass. The environmental
    density $\rhoR$ and redshift $z$ provide $\sim5\times$ smaller
    average contributions. The
    environmental temperature $\TR$ and nearby galaxy mass ratio $\upR$
    provide a relative contribution to stellar mass roughly $10\times$
    smaller than $\mvir$.
    Panel b) shows the
    feature functions contributing to the
    CEBM $\cebmms$. The stellar
    mass increases
    with $\mvir$, which provides the largest contribution. Secondary
    contributions come from environmental density $\rhoR$ and redshift $z$.
    Environmental
    temperature $\TR$ has a smaller contribution, but shows the familiar
    feature at $\TR\approx10^{4}$K where hydrogen ionizes. The mass
    ratio of nearby halos $\upR$ provides a minor contribution.
    As expected, the CEBM feature functions resemble the base EBM
    feature functions that represent the parameter dependence of stellar mass
    for most galaxies in the dataset (see Panel b) of Figure \ref{fig:mstar_base_ebm}).
    Panel c) presents the
    interaction functions for the CEBM $\cebmms$. Each panel
    shows the contribution of the interaction terms, normalized
    such that the color map ranges between plus or minus the maximum of
    the norm of each function $||f||_\mathrm{max}$. Teal indicates
    negative contributions and green indicates positive
    contributions.
    The table lists
    $||f||_\mathrm{max}$ for the interaction functions, each with
    units $\log_{10}\Msun$. As for the interaction functions
    for the base EBM $\basems$, the largest
    interaction occurs for large virial mass $\mvir$ and large environmental
    temperature $\TR$.
    Stellar mass is partially reduced for low environmental
    temperature $\TR$ and high environmental density $\rhoR$.
    These trends are similar to those for the base EBM $\basems$ modeling
    stellar mass (see Panel c) of Figure \ref{fig:mstar_base_ebm}).
    }
    \label{fig:mstar_composite_ebm}
\end{figure}

\begin{table}
    \centering
    \begin{tabular}{l c}
    \multicolumn{2}{c}{Average Contributions for the CEBM $\cebmms$} \\
    \toprule
    Feature & Value $[\log_{10} \Msun]$\\
    \midrule
    $\beta_{\log_{10}\mstar}$                           & $6.6995$ \\
    $\cfeat(\log_{10} \mvir)$                          & $0.5008$ \\
    $\cfeat(z)$                                        & $0.0961$ \\
    $\cfeat(\log_{10} \rhoR)$                        & $0.0902$ \\
    $\cfeat(\log_{10} \TR)$                          & $0.0576$ \\
    $\cfeat(\log_{10} \upR)$                    & $0.0336$ \\
    $\cfeat(z, \log_{10} \rhoR)$                     & $0.0172$ \\
    $\cfeat(\log_{10} \mvir, \log_{10} \rhoR)$       & $0.0108$ \\
    \bottomrule
    \end{tabular}
    \caption{Summary of the CEBM model $\cebmms$ trained to predict $\mstar$ using
    the restricted parameter set $\vtheta'$.
    The first entry, $\beta_{\log_{10}\mstar}$, is
             the learned baseline of the model. The next seven entries are the
             learned functions with the highest average contribution in
             descending order. The average contribution is computed via
             Equation \ref{eqn:cebm_average_contribution} (see \S
             \ref{appendix:cebm_average_contribution} for more details).}
    \label{table:mstar_composite_ebm_overall}
\end{table}

\end{document}